\documentclass[12pt]{article}

\author{Sebastian Bervoets\thanks{Aix-Marseille University (Aix-Marseille School of Economics), CNRS and EHESS. Email: sebastian.bervoets@univ-amu.fr} $\,$ Mario Bravo\thanks{Universidad de Santiago de Chile, Departamento de Matem\'atica y Ciencia de la Computaci\'on. Email: mario.bravo.g@usach.cl} $\,$ 
and Mathieu Faure\thanks{Aix-Marseille University (Aix-Marseille School of Economics), CNRS and EHESS. Email: mathieu.faure@univ-amu.fr}}
\title{Learning with Minimal Information in Continuous Games\thanks{ We wish to thank Y. Bramoull\'e, P. Mertikopoulos, N. Allouch, F. Dero\"ian, M. Belhaj, and participants of the various seminars where we presented this paper. The authors also thank the French National Research Agency (ANR) for their financial support through the ANR programs SOSOSO (13 JSH1 0009 01) and CIGNE (ANR-15-CE38-0007-01). Mario Bravo  gratefully acknowledges the support provided by FONDECYT 11151003.}
\\~
}
  
\def\hop{{\noindent}}
\newtheorem{theorem}{Theorem}

\newtheorem{hypothesis}{Hypothesis}
\newtheorem{lemma}{Lemma}
\newtheorem{proposition}{Proposition}
\newtheorem{definition}{Definition}
\newtheorem{example}{Example}

\newtheorem{remark}{Remark}

\newcommand{\norm}[1]{\left\Vert #1 \right\Vert}

\RequirePackage{natbib}

\usepackage{graphicx}
\usepackage{tikz}
\usepackage{pgfplots}
 \usepackage{pgfplotstable}
\usepackage[left=3cm,right=3cm,top=2.5cm,bottom=3cm]{geometry}
\usepackage{amssymb}
\usepackage{amsmath}
\usepackage{caption}
\usepackage[normalem]{ulem}

\usepackage{chngcntr}
\usepackage{apptools}
\AtAppendix{\counterwithin{lemma}{section}}
\AtAppendix{\counterwithin{theorem}{section}}
\AtAppendix{\counterwithin{proposition}{section}}
\AtAppendix{\counterwithin{example}{section}}
\AtAppendix{\counterwithin{definition}{section}}
\DeclareMathOperator{\inter}{Int}
\DeclareMathOperator{\sgn}{sgn}
\DeclareMathOperator{\BR}{BR}
\DeclareMathOperator{\Sp}{Sp}
\DeclareMathOperator{\Rp}{Re}
\DeclareMathOperator{\argmax}{argmax}

\begin{document}
\maketitle
\setlength{\baselineskip}{1\baselineskip}

\begin{abstract}
We introduce a stochastic learning process called the {\it dampened gradient approximation process}. While learning models have almost exclusively focused on finite games, in this paper we design a learning process for games with continuous action sets. It is payoff-based and thus requires from players no sophistication and no knowledge of the game. We show that despite such limited information, players will converge to Nash in large classes of games. In particular, convergence to a Nash equilibrium which is stable is guaranteed in all games with strategic complements as well as in concave games; convergence to Nash often occurs in all locally ordinal potential games; convergence to a stable Nash occurs with positive probability in all games with isolated equilibria.  
\end{abstract} 

\section{Introduction}

In this paper we construct a simple stochastic learning rule with the three following properties: (i) it is designed for games with continuous action sets; (ii) it requires no sophistication from the players and (iii) it converges to Nash equilibria in large classes of games. 
The question of convergence to Nash equilibria by agents playing a game repeatedly has given rise to a large body of literature on learning. One branch of this literature explores whether there are learning rules - deterministic or stochastic - which would converge to Nash equilibria in any game (see i.e. \cite{HartMas2003uncoupled}, \cite{HartMas2006}, \cite{babichenko2012completely}, \cite{foster2006regret}, Germano and Lugosi\cite{germano2007global}). Another branch, to which this paper contributes, focuses on specific learning rules and on the understanding of their asymptotic behavior. 

Both branches have almost exclusively addressed the issue of learning in discrete games (i.e. games where the set of strategies is finite). However, many economic variables such as price, effort, time allocation,  are non-negative real numbers, and thus are continuous. Typical learning models that have been designed for discrete games cannot be adapted to continuous settings without major complications, since they usually rely on assigning a positive probability to choosing each action, which cannot be done in continuous games. In this paper we introduce a learning rule designed for continuous games, which we call the {\it dampened gradient approximation process} (DGAP), and we analyze its behavior in several well-known classes of games. 

Learning rules can be more or less demanding in terms of players' sophistication and of the amount of information required to implement them. The DGAP belongs to the category of so-called {\it payoff-based} or {\it completely uncoupled} learning rules, meaning that players know nothing about the payoff functions (neither theirs nor those of their opponents), and they know nothing about the other players' actions, nor about their payoffs. They may not even be aware that they are playing a game. They only observe their own realized payoffs after each iteration of the game and make decisions based on these observations. 

Agents aim at maximizing their payoffs by choosing an action. If players knew the gradient of their utility function at every point, a natural learning process in continuous games would be for agents to follow a gradient method (see for instance \cite{arrow1960stability}). However, because players neither know the payoff functions nor observe the others' actions, they would be unable to compute these gradients. 

In DGAP, agents construct an approximation of the gradient at the current action profile,  
by randomly exploring the effects of increasing or decreasing their actions by small increments. The agents use the information collected from this exploration to choose a new action: if the effect revealed is an increase (resp. decrease) in payoff, then players move in the same (resp. opposite) direction, with an amplitude proportional to the approximated gradient. 

Although this procedure resembles a gradient learning process, there are two major differences from a standard gradient method. First, the DGAP is a random process instead of a deterministic dynamical system.  Second, in the standard gradient method with non-negative actions, players' behaviors are discontinuous at the boundary of the strategy space (see \cite{arrow1960stability}). In order to avoid such discontinuity in players' behavior, we assume that changes in actions are dampened as they approach the boundary. Hence the name of our learning process.  

We first prove that this process is well-defined - i.e. players' actions always remain non-negative (Proposition \ref{pr:stochapp}). Then we analyze its convergence properties and find that contrary to discrete games, where convergence to Nash of specific learning processes is generally difficult to obtain even for two- or three-player games, convergence is obtained in large classes of games with arbitrary numbers of players. We restrict to strongly single-peaked payoff functions\footnote{See Hypothesis \ref{hyp:conc} for a proper definition.}, focusing our attention on three classes of games that are of particular interest for economics and have been extensively analyzed in the learning literature: games with strategic complements, a class of games containing all potential games, and all games where the set of Nash equilibria is finite. This last class includes all games with a unique Nash equilibrium, such as strictly concave games and many of the generalized continuous zero-sum games. 

The DGAP is a stochastic process, the random part being the direction chosen for the exploration. We analyze its (random) set of accumulation points, called the {\it limit set}
, by resorting to {\it stochastic approximation theory}. This theory tells us that the long-run behavior of the stochastic process is driven by some underlying {\it deterministic dynamical system}. We thus start by showing that the deterministic system that underlies our specific stochastic learning process is a dampened gradient system (Proposition \ref{pr:stochapp}). We also show that all the Nash equilibria of a game are stationary points - otherwise called zeros - of this dynamical system, although other points may also be stationary. However, we prove (Proposition \ref{pr:OZ_unst}) that non-Nash stationary points are necessarily unstable\footnote{Throughout the paper, several notions of stability will be used. They are all recalled in Section \ref{model}.}. 
This is done in Section \ref{model}, where we also detail the DGAP and provide the necessary definitions. 

Stochastic approximation theory tells us that the stationary points of the underlying dynamical system are plausible candidates for the limit set of the random process. Yet it does not provide general criteria for excluding some of these candidates so as to obtain more precise predictions. This is actually one of the major difficulties in the field (see for instance \cite{BenFau12}). While the conceptual contribution of this paper lies in providing a natural learning process for games with continuous action sets, our technical contribution lies in providing precise statements on the structure of the limit set of the DGAP. Each result that we get is different, in the sense that it uses a different mathematical tool. It is remarkable that almost all our results hold with probability one, which is in general very difficult to obtain. 

In Section \ref{ssec:complements}, we analyze {\it games with strategic complements}. We show (Theorem \ref{th:convcompl}) that the DGAP cannot converge\footnote{Because the process is stochastic, the notion of convergence we use here is that of almost sure convergence.} to an unstable Nash equilibrium. 
Furthermore, we prove that the process will almost surely converge to a Nash equilibrium which is stable, except in very specific cases involving the structure of interactions between players: non-convergence might occur under a condition called {\it bipartiteness}. 

In Section \ref{ssec:pot}, we analyze a class of games that we call {\it locally ordinal potential games}. This class contains all the ordinal potential games, which in turn contain all the potential games. We have three results (in Theorems \ref{th:pot_nash} and \ref{th:attractor}). First, the limit set of the DGAP is always contained in the set of stationary points of the dynamics. When equilibria are isolated, this implies that the process converges to a Nash equilibrium with probability one, since we prove that the process cannot converge to a non-Nash stationary point in these games. Second, we show that under the condition of non-bipartiteness, the DGAP converges to a Nash equilibrium which is stable when equilibria are isolated. Third, 
although convergence to unstable stationary points (possibly non-Nash) cannot be ruled out in general (i.e. when equilibria are not isolated), we characterize the set of stable stationary points. We prove that they are local maxima of the locally ordinal potential function, that they are necessarily connected components of Nash equilibria, and that they are necessarily stable equilibria of another, unrelated dynamical system: the Best-Response dynamics. 

Finally, in Section \ref{ssec:isolated}, we consider all games for which stationary points are isolated. This class includes the vast majority of games studied in economics. We cannot prove precise and general convergence results, since there is no guarantee that the limit set of the process will be included in the set of stationary points. Still, we state two results. First, DGAP will converge to a stable Nash equilibrium with positive probability in all these games. Second, we exclude convergence to what we call undesirable stationary points, i.e. those that are non-Nash, and unstable Nash equilibria. 

Also in Section \ref{ssec:isolated}, we focus on games with a unique Nash equilibrium that have previously been analyzed, either by \cite{arrow1960stability} or by \cite{rosen1965existence}. They examined dynamical systems which are either discontinuous or complex gradient systems, obtaining convergence to the unique Nash equilibrium. We obtain the same results - for our process - with convergence to the unique equilibrium with probability one. Thus, another contribution of this paper is to show that we can use a gradient system which is both simple and continuous,
 and preserve the
convergence properties. 

\vspace{.3cm}

\noindent {\it \bfseries Related Literature}
\vspace{.1cm}

As mentioned earlier, the learning literature has essentially focused on finite action games. Many rules have been proposed and studied, both in the non payoff-based and in the payoff-based contexts. In the former, the most widely-explored adaptive process is fictitious play (introduced in \cite{Bro51}), where players' average actions are shown to converge\footnote{In all of the following papers, the convergence notions differ. This has implications in terms of the scope of each result. For details the reader should refer directly to the papers.} for $2$-player zero-sum games (\cite{Rob51}), for $2\times N$ games (\cite{Miy61} and \cite{Ber05}), for potential games (\cite{MonSha96b}). Convergence is also obtained for {\it stochastic fictitious play}, introduced by \cite{FudKre93}, in $2\times 2$ games (\cite{BenHir99}), zero-sum and potential games (\cite{HofSan02}) and supermodular games (Bena{\"\i}m and Faure\cite{BenFau12}). However, it has been shown that fictitious play does not always converge to Nash once there  are at least $3$ actions per player (\cite{Sha64}). Other non payoff-based learning rules include hypothesis testing (\cite{foster2003learning}) or calibrated forecasting (\cite{kakade2008deterministic}). Our contribution differs in both dimensions: we focus on continuous games and on payoff-based procedures. 

Many payoff-based learning rules have been explored in the context of discrete games. Such rules include the popular class of reinforcement learning procedures (see \cite{BorSar97} or \cite{EreRot98} for pioneer work). These procedures have been studied in very specific finite games: $2 \times 2$ games in \cite{Pos97}, $2$-player games with positive payoffs in \cite{BorSar97}, \cite{Beg05}, \cite{HopPos05} or \cite{Las01}. On the same topic, see also \cite{LesCol05}, \cite{CMS10}, \cite{BraFau15} and \cite{Bravo16}. However, it is not known if these procedures converge to Nash in more general games.

Other payoff-based procedures for discrete games have been proposed, including: Regret-testing (\cite{foster2006regret}) which converges in any $2$-person game; Generalized regret-testing (\cite{germano2007global}) which converges in any generic $n$-person game; Experimentation dynamics (\cite{marden2009payoff}) which converge to Nash in the class of $n$-person weakly acyclic games; Trial and error (\cite{young2009learning}) which comes close to Nash equilibrium a large fraction of the time; Aspirational learning (\cite{karandikar1998evolving}) which may fail to converge even in $2\times 2$ games.

The literature on continuous game is sparser, and a distinction can also be made between procedures which are demanding in terms of sophistication and knowledge of the players, and procedures which are of the payoff-based type. In the first category, \cite{arrow1960stability} prove that when all players' payoff functions are strictly concave, the gradient method converges to the unique Nash equilibrium in generalized zero-sum games. \cite{rosen1965existence} studies a gradient method in concave $n$-person games with a unique equilibrium, and shows that this unique equilibrium is globally asymptotically stable for some weighted gradient system, with suitably chosen weights. In a recent paper, \cite{Merti2018} studies a gradient-like stochastic learning algorithm where agents receive erroneous information about their gradients in the context of concave games, and shows that whenever this process converges, it does so to a Nash equilibrium. Using a different approach, \cite{LesPer14} adapt stochastic fictitious play to games with continuous action sets, and show that it converges in $2$-player zero-sum games. Our contribution differs from these in analyzing a payoff-based learning process.

The two papers most closely related to ours are those analyzing payoff-based procedures designed for continuous games. \cite{DindosM06} consider a stochastic adjustment process called the better-reply. At each step, agents are sequentially picked to play a strategy chosen at random, while the other players do not move. The agent then observes the hypothetical payoff that this action would yield, and decides whether to stick to this new strategy or to go back to the previous one. This process converges to Nash when actions are either substitutes or complements around the equilibrium in games called aggregative games, with quasi-concave utility functions. However, their contribution differs from ours in several respects, the most important being that agents revise their strategies sequentially. The driving force for convergence is that with positive probability, every player will be randomly drawn as many times as necessary to approximate a best response. In our paper, we assume that all players move simultaneously. In that case, it is easy to construct a simple game where simultaneity drives the better-reply adjustment process to cycle. 

The second related paper, by \cite{huck2004through}, considers another type of payoff-based learning process, called Trial and error - but which has no link with the \cite{young2009learning} procedure - in the context of the Cournot oligopoly, where players move simultaneously, as in our paper. Players choose a direction of change and stick to this direction as long as their payoff increases, changing as soon as it decreases. The authors show that the process converges, but it does so to the joint-profit maximizing profile and not to the (unique) Nash equilibrium of the game. While our paper is similar to theirs in spirit, unlike them, we do not focus on a specific game\footnote{Although, as the authors suggest, the intuition for their result might carry through to other games.}, we allow for multiple equilibria, including continua of equilibria, and we get convergence to Nash. Our process differs in two respects. First, we consider an exploration stage where the players decide in which direction they will be moving. This exploration stage seems to make our agents less naive. Second, the amplitude of the moves is constant in their paper while in ours, amplitude depends on the variation in payoffs after an exploration stage. This might explain why we obtain convergence to Nash while they do not.

A group of papers coming from a different literature deserves mention here: in the literature on evolutionary dynamics in population games, many dynamical systems have been suggested in the context of infinite populations choosing strategies among a finite set. Recently, several papers have extended these dynamical systems to continuous strategy spaces. The continuous strategy version of the replicator dynamic has been studied by \cite{bomze1990dynamical}, \cite{oechssler2001evolutionary}, \cite{oechssler2002dynamic}, and \cite{cressman2005stability}, while the Brown-von Neumann-Nash (BNN) dynamic has been extended by \cite{hofbauer2009brown}. \cite{lahkar2015logit} extend the  logit dynamics, \cite{cheung2014pairwise} adapts the pairwise comparison dynamics and
\cite{cheung2016imitative} works with the class of imitative dynamics (which includes the replicator dynamics). Although this group of papers deals with dynamical systems for continuous games, their contexts are totally different (continuum of players). Moreover, their main goal is to define the extensions of existing dynamics, and to see whether they are well-defined and share the same properties as their discrete-strategy counterparts.

\section{The model}\label{model}

\subsection{Definitions and hypothesis}

Let $\mathcal{N} = \{1, \ldots, N\}$ be a set of players, each of whom repeatedly chooses an action from $X_i = [0,+\infty[$. An action $x_i \in X_i$ can be thought of as an effort level chosen by individuals, a price set by a firm, a monetary contribution to a public good, etc. Let $X =  \times_{i=1,...,N} X_i$. We denote by 
$\partial X$ the boundary of $X$, i.e. $\partial X := \{x \in X; x_i = 0 \text{ for some } i \in \mathcal{N}\}$. And we let $\inter(X):= X \setminus \partial X$ denote the interior of $X$.

At each period of time, players observe a payoff that is generated by an underlying repeated game $\mathcal{G} = (\mathcal{N},X,u)$, where $u = (u_i)_{i=1,...,N}$ is the vector of payoff functions. Players know nothing about the payoff functions, nor about the set of opponents. In this paper we will examine several classes of underlying games, each class being defined by different properties on the functions $u_i$. However, we will always make the two following standing assumptions: 

\begin{hypothesis} \label{hyp:conc}
For any $i$, the payoff map $u_i$ is assumed to be $\mathcal{C}^1$ on $\mathbb{R}_+^N$ and with the property that, for any $x_{-i} \in X_{-i}$, there exists $M(x_{-i}) \in X_i$ such that the map $x_i \mapsto \frac{\partial u_i}{\partial x_i}(x_i,x_{-i})$ is strictly positive for $x_i < M(x_{-i})$ and strictly negative for $x_i >  M(x_{-i})$.
\end{hypothesis} 
Hypothesis \ref{hyp:conc} implies that best responses are unique and $\BR_i(x_{-i}) = M(x_{-i})$. This assumption is verified for instance if $x_i \mapsto u_i(x_i,x_{-i})$ is strictly concave, $\frac{\partial u_i}{\partial x_i}(0,x_{-i}) >0$ and $\lim_{x_i \rightarrow + \infty} \frac{\partial u_i}{\partial x_i}(x_i,x_{-i}) < 0$. 

In the games we consider, interactions between players can be very general. They can be heterogeneous across players and they can be of any sign. However we assume that externalities are symmetric in sign:

\begin{hypothesis} \label{hyp:sym_ext} Games are assumed to have {\it symmetric externalities}, i.e. $\forall i \neq j$ and $\forall x$, 
\begin{eqnarray*} \label{eq:sym_ext}
\sgn \left(\frac{\partial u_i}{\partial x_j}(x) \right) = \sgn \left(\frac{\partial u_j}{\partial x_i}(x) \right)
\end{eqnarray*} where 
$\sgn (a) = 0$ if   $a=0$.
\end{hypothesis}

Most of the continuous games in the economics literature fall into this class. Note that a game with symmetric externalities does not require them to be of equal intensity.
Also, symmetric externalities allow for patterns where $i$ exerts a positive externality on individual $j$ and a negative externality on individual $k$. 
Note finally that symmetric externalities do not imply that $\sgn \left(\frac{\partial u_i}{\partial x_j}(x) \right) = \sgn \left(\frac{\partial u_i}{\partial x_j}(x') \right)$ for $x \neq x'$. 

Some of our results will depend on the pattern of interactions in the game $\mathcal{G}$. We capture this pattern by an {\it interaction graph}, defined as follows. Let $x = (x_1, \ldots, x_N)$ be an action profile.  The interaction graph at profile $x$ is given by the matrix $\mathbf{G}(x)$ where $g_{ii}(x) =0$ and, for $i \neq j$, $g_{ij}(x) = 1$ if $\frac{\partial u_i}{\partial x_j}(x) \neq 0$ and $g_{ij}(x) = 0$ otherwise. Note that the interaction graph is local, in the sense that it depends on the vector of actions. Thus $\mathbf{G}(x)$ can either be constant on $X$ or change as $x$ changes. Note also that the interaction graph of a game satisfying Hypothesis \ref{hyp:sym_ext} is symmetric. 

We now provide two examples of games satisfying Hypothesis \ref{hyp:conc} and \ref{hyp:sym_ext} and describe the interaction graphs.

\begin{example}[Public good game]
Players contribute an effort $x \in [0; +\infty[$ to a public good. The payoff of player $i$ is $u_i(x) = b_i(x_i +  \sum_{j\neq i} \delta_{ij} x_j) - c_i.x_i$, where $c_i > 0$ is the marginal cost of effort for $i$, $1 \geq \delta_{ij} \geq 0$ is a measure of subtitutability between $i$ and $j$'s efforts, and $b_i(.)$ is a differentiable, strictly increasing concave function. In some contexts, not all players will benefit from one player's contribution (see \cite{BraKra07}). This is why we leave the possibility that $\delta_{ij} = 0$ for some pairs $(i,j)$. 

This game satisfies Hypothesis \ref{hyp:conc} and \ref{hyp:sym_ext}. Further, the interaction graph is constant since it does not depend on the action profile $x$: $g_{ij}(x) = 0 \text{ if } \delta_{ij} = 0$, while $g_{ij}(x) = 1 \text{ if } \delta_{ij} > 0$.
\end{example}

\begin{example}[Aggregate demand externalities]
In macro-economics, aggregate-\phantom{  }\linebreak demand-externality models (see for instance \cite{fudenberg1991game}) are games satisfying Hypothesis \ref{hyp:conc} and \ref{hyp:sym_ext}. For instance, search models \`a la Diamond enter this class, where players exert a search effort $x_i$, with payoff functions $u_i(x) = \alpha x_i \sum_{j\neq i} x_j -c(x_i)$, and where $c(x_i)$ is the cost of searching, $x_i x_j$ is the probability that $i$ and $j$ end up partners and $\alpha$ is the gain when a partner is found. The interaction graph of such a game depends on the profile $x$, since $\frac{\partial u_i}{\partial x_j}(x) = \alpha x_i$, thus $g_{ij}(x) = 0 \text{ if } x_i = 0$, while $g_{ij}(x) = 1 \text{ if } x_i \neq 0$.

This game can be generalized to local search models with payoffs  $u_i(x) = \alpha x_i \sum_{j\neq i} \delta_{ij} x_j -c(x_i)$, where, as above, $\delta_{ij} \geq 0$ and where $\delta_{ij}=0$ means that individual $j$ exerts no externality on individual $i$. A popular game analyzed in the network literature in a different context is the game introduced in \cite{BCZ06}, which is actually a local aggregate-demand-externality model, where $c(x_i) = \frac{1}{2} x_i^2 - x_i$, and $\delta_{ij} \in \{0,1\}$. In that case, the interaction graph only partly depends on $x$, since $\delta_{ij} = 0$ implies $g_{ij}(x) = 0$ for all $x$.  
\end{example}

We denote by $NE$ the set of Nash equilibria. In many economics applications, Nash equilibria would consist of isolated points. Examples 1 and 2 provide examples of games where the set of Nash equilibria is generically finite. However, in what follows we will sometimes deal with a continuum of equilibria. This is the case for instance in Example 1, when $\delta_{ij} \in \{0, 1\}$ (see \cite{BervFaure}). Because we wish to be as general as possible, we will consider \emph{connected components} of $NE$:

\begin{definition} Let $\Lambda$ be a compact connected subset of $NE$ and let $N^{\delta}(\Lambda) := \{y \in X: d(y,\Lambda) < \delta\}$. 
We say that $\Lambda$ is a \emph{connected component} of $NE$ if  there exists $\delta >0$ such that $N^{\delta}(\Lambda) \cap NE = \Lambda$. 
\end{definition}

\subsection{The Learning Process}
\vspace{.3cm}

We consider a payoff-based learning process in which agents construct a partial approximation of the gradient by exploring the effects of deviating in one direction that they chose at random at every period. This information allows agents to choose a new action depending on what they just learned from the exploration stage. 
Here we detail what agent $i$ does, bearing in mind that every agent simultaneously uses the same rule.
\vspace{.4cm}

{\tiny $\bullet$} At the beginning of round $n$, agent $i$ is playing action $x_n^i := e_{2n}^i$ and is enjoying the associated payoff $u_i(e_{2n}^i, e_{2n}^{-i})$. Player $i$ then selects his actions $e_{2n+1}^i$ and $e_{2n+2}^i (= x_{n+1}^i)$ as follows.

{\tiny $\bullet$} {\it Exploration stage -} Player $i$ plays a new action $e_{2n+1}^i$, chosen at random around his current action $e_{2n}^i$. Formally, let $(\epsilon^i_n)_n$ be a sequence of i.i.d random variables such that $\mathbb{P}(\epsilon_n^i=1) =  \mathbb{P}(\epsilon_n^i=-1)=1/2$. At period $n$, $\epsilon_{n}^i$ is drawn and player $i$ plays $e^i_{2n+1} := e^i_{2n}  + \frac{1}{n+1} \epsilon_{n}^i$. 

{\tiny $\bullet$} {\it Updating stage -} Player $i$ observes his new payoff, and computes 
\vspace{-1ex}
$$\Delta u_{n+1}^i := u_i(e^i_{2_n+1},e^{-i}_{2_n+1}) - u_i(e^i_{2n},e^{-i}_{2n}).$$
 This quantity provides $i$ with an approximation of his payoff function's gradient. Using this information, player $i$ updates his action by playing $e^i_{2n+2}:= e^i_{2n} + \epsilon^i_n \Delta u_{n+1}^i e^i_{2n}$. Thus, when $\Delta u_{n+1}^i$ is positive, player $i$ follows the direction that he just explored, while he goes in the opposite direction when $\Delta u_{n+1}^i$ is negative.

{\tiny $\bullet$} Period $n$ ends. We set $x_{n+1}^i := e_{2n+2}^i$ and agent $i$ gets the payoff $u_i(e^i_{2n+2}, e^{-i}_{2n+2})$. Round $n+1$ starts. 
\vspace{.3cm}

Let $x_n = e_{2n}$ and $\mathcal{F}_n$ be the history generated by $\{e_1,...,e_{2n+1}\}$. Studying the asymptotic behavior of the random sequence $(e_n)_n$ amounts to studying the sequence $(x_n)_n$. Hence the focus of this paper is on the convergence of the random process $(x_n)_n$.

The next proposition shows that the process is well-defined, in the sense that it always remains within the admissible region (i.e. the actions stay positive). It also proves that the DGAP is a discrete time stochastic approximation process. 

\begin{proposition} \label{pr:stochapp}
The iterative process is such that $x_n^i > 0$ for all $i$. \\
It can be written as 
\begin{equation}\label{ode1}
x_{n+1} = x_{n} +\frac{1}{n+1} \left(F(x_{n}) + U_{n+1}+ \xi_{n+1}  \right),
\end{equation}
where
\begin{itemize}
\item[$(i)$] $F(x) = (F_i(x))_i$ with $F_i(x) = x_i \frac{\partial u_i}{\partial x_i}(x_i,x_{-i})$,
\item[$(ii)$] $U_{n+1}$ is a bounded martingale difference (i.e  $\mathbb{E}\left(U_{n+1} \mid \mathcal{F}_n \right) = 0$),
\item[$(iii)$] $\xi_n = \mathcal{O}(1/n)$.
\end{itemize}
\end{proposition}

\vspace{.2cm} 
All our proofs are in the appendix.

The iterative process \eqref{ode1} is a discrete time stochastic process with step $\frac{1}{n+1}$.\footnote{Note that $\sum_n \frac{1}{n+1} = \infty$ and $\lim_{n \rightarrow \infty} \frac{1}{n+1} = 0$. It is important that the sum diverges, to guarantee that the process does not get "stuck" anywhere, unless agents want to stay where they are. Further, it is important that the terms go to zero, so that the process can "settle" when agents want to. In fact, the term $\frac{1}{n+1}$ can be replaced by any step of the form $\alpha_{n+1}$, where $\sum_n \alpha_{n+1} = \infty$ and $\lim_{n \rightarrow \infty} \alpha_{n+1} = 0$, without affecting the results.}. If there were no stochastic term, the process \eqref{ode1} would write
\begin{equation*}\label{euler}
x_{n+1} = x_{n} +\frac{1}{n+1} F(x_{n}),
\end{equation*}
which corresponds to the well-known Euler method, a numerical procedure for approximating the solutions of the deterministic ordinary differential equation (ODE)

\begin{equation}\label{edo}
\dot{x} = F(x).
\end{equation} 

Although the (stochastic) process \eqref{ode1} differs from the (deterministic) process \eqref{edo} because of the random noise,  the asymptotic behavior of \eqref{edo} will inform us on the asymptotic behavior of \eqref{ode1}.\footnote{Stochastic approximation theory (see \cite{Ben96} or \cite{Ben99} for instance) tells us that, as periods unfold, the random process  gets arbitrarily close to the solution curve of its underlying dynamical system. In other words, given a time horizon $T>0$ - however large it might be - the process shadows the trajectory of some solution curve between times $t$ and $t+T$ with arbitrary accuracy, provided $t$ is large enough.}

\begin{remark}\label{rk:rk1} In the standard gradient method (see for instance \cite{arrow1960stability}), the dynamical system is defined as $\dot{x} = H(x)$ where 
\begin{equation}\label{Syst_Arrow}
H_i(x) =
  \left\{
      \begin{array}{ll}
        \frac{\partial u_i}{\partial x_i}(x) & \text{ unless } x_i = 0 \text{ and } \frac{\partial u_i}{\partial x_i}(x) < 0,\\
        0  & \text{ otherwise}.   
      \end{array}
    \right.
\end{equation}
The function $H$ is thus discontinuous, since the process could otherwise leave the admissible space. Conversely, the dynamical system underlying the DGAP is continuous: $F_i(x) = x_i \frac{\partial u_i}{\partial x_i}(x)$. The role played by the multiplicative factor $x_i$ is to dampen the variations of the state variable and ensure that it will never reach the boundary - although it can converge to it. This is not unreasonable behavior: the gradient system assumes that players crash onto the boundary, whereas we assume that the closer they get to the boundary, the smaller their movements become. 

In \cite{rosen1965existence}, the author studies another gradient method also ensuring that the system never leaves the state space. The system is given by
\begin{equation}\label{Syst_Rosen}
H_i(x) = r_i \frac{\partial u_i}{\partial x_i}(x) + \sum_{j=1}^{k} \lambda_j \frac{\partial h_j}{\partial x_i}(x),
\end{equation}
where the $h_j$ functions are the constraints defining the convex and compact set $R$ where $x$ lives (i.e. $R = \{x; h(x) \geq 0\}$), $r_i$ and $\lambda_j$ are appropriately chosen weights guaranteeing that the system will always remain within the set $R$. 
\end{remark}

\subsection{Limit sets} \label{ssec:lim_sets}

The focus of this paper is on the asymptotic behavior of the random process $(x_n)_n$. 
Hence we are interested in its \emph{limit set}\footnote{In the remainder of the paper, we will always place ourselves on the event $\{\limsup_n \|x_n\| < + \infty \}$, i.e. we will abstract from the possible realizations which take the process to infinity.}. 

\begin{definition} [Limit set of $(x_n)_n$] Given a realization of the random process, we denote the limit set of $(x_n)_n$ by 
\begin{equation}
\mathcal{L}\left((x_n)_n \right):= \{x \in X ; \; \exists \mbox{ a subsequence $x_{n_k}$ such that} ; \; x_{n_k} \rightarrow x\}.
\end{equation} 
\end{definition}

Note that the limit set of the learning process is a random object, because the asymptotic behavior of the sequence $(x_n)_n$ depends on the realization of the random sequence $(\epsilon_n)_n$, drawn at every exploration stage.   

Proposition \ref{pr:stochapp} allows us to make use of stochastic approximation theory, which provides a characterization of the candidates for $\mathcal{L}\left((x_n)_n \right)$\footnote{In Bena\"{i}m (1999), it is established that on \mbox{$\{\limsup_n \|x_n\| < + \infty\}$}, the limit set of $(x_n)$ is always compact, invariant and attractor-free. This class of sets is called \emph{internally chain transitive} (ICT). These sets can take very complicated forms, but they conveniently include the zeroes of $F$ and the $\omega$-limit set of any point $x$ (if non-empty).}.
In particular the $\omega$-limit sets of $\dot{x}= F(x)$\footnote{Let $\varphi(x,t)$ denote the flow of $F(.)$, i.e. the position of the solution of \eqref{edo} with initial condition $x$, at time $t$. Then, the $\omega$-limit set of $x$ is given by $\omega(x) := \{z \in X; \; \, lim_{k \rightarrow \infty} \varphi(x, t_k) = z \text{ for some } t_k \rightarrow \infty \}. $ 
Notice that by the regularity assumption on $u(.)$, $F$ satisfies the Cauchy-Lipschitz condition that guarantees that, for all $x \in X$, $\varphi$ is well-defined and unique. We consider the restriction of $\varphi$ on $X (= \mathbb{R}^N_+)$, since $X = \mathbb{R}^N_+$ is invariant for its flow, and our random process \eqref{ode1} always remains in the positive orthant. } lie among these candidates.   However, several difficulties remain: first, there might be other candidates that are not $\omega$-limit sets of the underlying ODE. Moreover, this theory does not provide general criteria to systematically exclude any of these candidates, nor to confirm that they are indeed equal to $\mathcal{L}\left((x_n)_n \right)$.


The stationary points of the dynamical system \eqref{edo} are particular $\omega$-limit sets that will be of interest to us, as they contain all the Nash equilibria of the underlying game. The set of stationary points, denoted $Z(F)$, will be called the {\it zeros of $F$}: $Z(F):= \{x \in X; F(x) = 0\}$. For convenience, we drop the reference to $F$ and simply write $Z$. 

Observe that $F_i(x) = x_i \frac{\partial{u_i}}{\partial x_i}(x)$. Thus, 
$x \in Z \Longleftrightarrow \forall i \in \mathcal{N}, \left(x_i = 0 \; \text{ or } \; \frac{\partial u_i}{\partial x_i}(x_i,x_{-i}) = 0 \right)$,
while 
$x \in NE \Longleftrightarrow \; \, \forall i \in \mathcal{N}, \left(\frac{\partial u_i}{\partial x_i}(x_i,x_{-i}) = 0, \text{ or } \; x_i=0 \text{ and } \;  \frac{\partial u_i}{\partial x_i}(x_i,x_{-i}) \leq 0 \right) $.
This implies that all the Nash equilibria of the game are {\it included} in the set of zeros of $F$. Unfortunately, $Z$ contains more than the set of Nash equilibria.  We call $x \in Z \setminus NE$ an {\it other zero} ($OZ$) of the dynamical system: 
\vspace{-1ex}
$$OZ = \left\{x: \; F(x)=0 \text{ and } \exists i \, \text{ s.t.} \; x_i=0, \;  \frac{\partial u_i}{\partial x_i}(x) >0\right\}.$$
\noindent We have the following partition of F:
\begin{equation} \label{eq:OZ}
Z = NE \cup OZ.
\end{equation}
Note that $\partial X$ might contain some points in $NE$, however $OZ \subset \partial X$.\\


Convergence or non-convergence of our random process to a given point or set will sometimes depend on the \emph{stability} of the latter with respect to the deterministic dynamical system $\dot{x}= F(x)$. In different sections we use various notions of stability, which we recall here. 

Let $\hat{x} \in Z$. The point $\hat{x}$ is \emph{asymptotically stable} (denoted by $\hat{x} \in Z^{AS}$) if it uniformly attracts an open neighborhood $W$ of itself: $\lim_{t \rightarrow + \infty} \sup_{x \in W} \|\varphi(x,t)-\hat{x}\|=0$, where $\varphi(x,t)$ denotes the flow of $F(\cdot)$.
The point $\hat{x}$ is \emph{linearly stable} (denoted by $\hat{x} \in Z^{LS}$) if for any $\lambda \in \Sp(DF(\hat{x}))$ - where $DF(\hat{x})$ is the Jacobian matrix of $F$ evaluated at $\hat{x}$ and $\Sp(M)$ is the spectrum of matrix $M$ - we have $\Rp(\lambda) <0$ and $\hat{x}$ is  {\it linearly unstable} (denoted by $\hat{x} \in Z^{LU}$) if there exists $\lambda \in \Sp(DF(\hat{x}))$ such that $\Rp(\lambda) >0$. 
Note that if $\hat{x}$ is hyperbolic (that is $\Rp(\lambda) \neq 0$ for any $\lambda \in \Sp(DF(\hat{x}))$) then it is either linearly stable or linearly unstable. We denote the set $Z \setminus Z^{LU}$ by $Z^S$ and by a slight abuse of language, we will call all points in $Z^S$ {\it stable}.

We have the following inclusions: 
\begin{equation*} \label{eq:inclusions} 
Z^{LS} \subset Z^{AS} \subset Z^S.
\end{equation*}

\begin{proposition} \label{pr:OZ_unst}
We have $OZ \subset Z^{LU}$. As a consequence, $Z^{S} \subset NE$.
\end{proposition}

To prove this, we take $x$ in $OZ$ and pick an individual such that $x_i = 0$ and $\frac{\partial u_i}{\partial x_i} > 0$. We then show that $\frac{\partial u_i}{\partial x_i}$ is an eigenvalue of $DF(x)$. 
The direct consequence of Proposition \ref{pr:OZ_unst} is that if the limit set $\mathcal{L}\left((x_n)_n \right)$ contains stable stationary points, they must be stable Nash equilibria.  

In view of Proposition \ref{pr:OZ_unst}, we will use the following notations in the remainder: $NE^{LS} := Z^{LS}$, $NE^{AS} := Z^{AS}$ and $ NE^S := Z^S$. \\
\\
As mentioned earlier, we will sometimes be dealing with connected components of $NE$ instead of isolated points.  We will thus use the concept of \emph{attractor} (see \cite{Rue81}):

\begin{definition} Let $S \subset X$ be invariant for the flow $\varphi$. Then a set $A \subset S$ is an attractor for $\dot{x} = F(x)$ if\\
$(i)$ $A$ is compact and invariant;\\
$(ii)$ there exists an open neighborhood $U$ of $A$ with the following property: 
\[\forall \epsilon >0, \; \, \exists\, T>0 \mbox{  such that }  \; \, \forall x \in U,\; \forall\, t \geq T, \; \,  d(\varphi(x,t),A) < \epsilon .\]
\end{definition}
An attractor for a dynamical system is a set with strong properties: it uniformly attracts a neighborhood of itself.

\begin{remark} Let $\hat{x} \in Z$ be an isolated stationary point of $\dot{x}=F(x)$. Then $\hat{x}$ is asymptotically stable if and only if $\{\hat{x}\}$ is an attractor for $\dot{x}=F(x)$. 
\end{remark}

We turn to the analysis of several classes of games. 

\section{Strategic complements}\label{ssec:complements}

\begin{definition}
A game $\mathcal{G}$ is a game with strategic complements if payoff functions are such that $\frac{\partial^2 u_i}{\partial x_i \partial x_j}(x) \geq 0$ for all $i \neq j$.

\end{definition}

Games with strategic complements have nice structured sets of Nash equilibria (\cite{Viv90}), and offer nice convergence properties for specific dynamical systems. However, it can be difficult to obtain convergence to Nash for general learning procedures. There are several reasons for this that we illustrate here through two examples. 

First, consider the Best-Response dynamics. Under Hypothesis \ref{hyp:conc}, best-response functions are differentiable and strictly increasing. In that case, \cite{Viv90} proves in Theorem 5.1 and Remark 5.2 that, except for a specific set of initial conditions, the Best-Response dynamics, whether in discrete or in continuous time, monotonically converges to an equilibrium point. Unfortunately, in our case this set of problematic initial conditions cannot be excluded, in particular because the process is stochastic. It could be that the stochastic process often passes through these points, in which case it is known to possibly converge to very complicated sets\footnote{See  \cite{Hir99}.}. In order to study convergence of the DGAP, we thus need to consider all possible trajectories and cannot rely on existing results. 

Second, consider the standard reinforcement learning stochastic process, whose mean dynamics are the replicator dynamics. As shown in \cite{Pos97}, the process can converge with positive probability to stationary points that are not only unstable, but also non-Nash. Examples can be constructed with $2$ players, each having $2$ strategies, supermodular payoff matrices with a unique strict Nash equilibrium, which is, moreover, found by elimination of dominated strategies. Yet even then, the learning process converges with positive probability to any other combination of strategies. This happens because there are some stationary points of the dynamics where the noise generated by the random process is null.

These two examples illustrate how, despite the games' appealing properties, convergence to Nash is neither guaranteed nor easy to show when it occurs. We show that the DGAP will converge. In order to get our result, we first need to prove that no point in $\partial X$, the boundary of the state space, will be included in the limit set of the process. We start by imposing a simple and natural hypothesis.

\begin{hypothesis} \label{hy:boundary} For any agent $i$, 
\[\frac{\partial u_i}{\partial x_i}(0,0) > 0.\]
\end{hypothesis}

Hypothesis \ref{hy:boundary} guarantees that players want to move away from the origin. Because of strategic complementarities, this also implies that players want to move away from any point of $\partial X$ (since $\frac{\partial u_i}{\partial x_i}(0,x_{-i}) > \frac{\partial u_i}{\partial x_i}(0,0)$). However, despite the fact that all players prefer to move away from the boundary, it is not clear why  the stochastic process should remain at a distance from this boundary. The difficulty comes from the following fact: assume players start close to the boundary. Then, at the exploration stage, some decrease their efforts while others increase theirs. Although complementarities imply that the players who decreased their efforts would have been better-off if they had instead increased them, they could still end up with a better payoff than before the exploration, and thus continue decreasing at the updating stage, getting closer to $\partial X$.

The following proposition proves that this will not happen in the long run.
\begin{proposition}\label{lm:nonconvboundary}
Under Hypothesis \ref{hy:boundary}, there exists $a >0$ such that $\mathcal{L}((x_n)_n) \subset [a, +\infty[^N$ almost surely.
\end{proposition} 
From the  mathematical point of view, the major problem to obtain Proposition \ref{lm:nonconvboundary} is to show that a stochastic approximation algorithm like the one given by \eqref{ode1} is pushed away from an invariant set for $F$ where the noise term vanishes. In fact, there is almost no general result along these lines in the literature. 

The proof of Proposition \ref{lm:nonconvboundary} is long and technical, but the idea goes as follows: among the players close to the boundary, the player exerting the least effort will increase his effort on average. Unfortunately, this does not imply that the smallest effort
also increases, since another player may have decreased his. We thus construct a stochastic process which is a suitable approximation of the smallest effort over time. We then show that this new process cannot get close to the boundary, and because it is close asymptotically to our process, we are able to conclude. 

\begin{definition}
The interaction graph $\mathbf{G}(x)$ is said to be bipartite at $x \in X$ if the set $N$ of players can be partitioned into $N_1$ and $N_2$ such that for any pair of players $i$ and $j$ we have 
\[ g_{ij}(x) = 1 \Longrightarrow \left(i \in N_1 \text{ and } j \in N_2 \right) \text{ or } \left(i \in N_2 \text{ and } j \in N_1\right). \]

An interaction graph is non-bipartite on a set $A$ if for all $x \in A$, $\mathbf{G}(x)$ is not bipartite.
\end{definition}

We are now ready to state the main result of this section.

\begin{theorem} \label{th:convcompl} Consider a game of strategic complements and smooth payoff functions, and assume that Hypothesis \ref{hy:boundary} holds. Then
\begin{itemize}
\item[(i)] The learning process $(x_n)$ cannot converge to an unstable Nash equilibrium:
\[\forall \, \tilde{x} \in NE^{LU},\; \mathbb{P} \left(\lim_n x_n = \tilde{x} \right) = 0.\]
\item[(ii)] If, in addition, the interaction graph is non-bipartite on $\inter(X)$, the learning process $(x_n)$ almost surely converges to a stable Nash equilibrium:
\[\mathbb{P} \left(\exists \, x^* \in NE^S: \; \, \lim_n x_n = x^* \right) = 1.\]
\end{itemize} 
\end{theorem}

This result is very tight. Because the hypotheses of the theorem are verified for most common economic models we can think of, this theorem guarantees that the learning process will not only converge to Nash in most cases, it will additionally converge to a stable equilibrium. 
In cases where the interaction graph is bipartite, we cannot guarantee that the process will not converge to general unstable sets\footnote{Linearly unstable equilibria are unstable sets, but unstable sets also include much more complex structures.}. However, we can still exclude convergence to linearly unstable equilibria by point $(i)$. 

Let us provide some insights on the bipartiteness condition. As in \cite{Pos97}, one potential issue is that the random process could get stuck around stationary points of the underlying dynamics if the random noise is zero at these stationary points. More precisely, a stationary point is unstable if there is some direction along which the system "escapes" the stationary point. But the system has to be able to follow that direction, otherwise it will get stuck. The random process plays precisely this role here: it allows the system to escape, as long as the unstable direction component of the random noise $(U_n)_n$ is not zero at that point. At an unstable equilibrium, we can show that the noise is not zero in the unstable direction and this guarantees the non-convergence result of part $(i)$. The non-bipartiteness of the network guarantees that the noise $(U_n)_n$ has the property of being {\it uniformly exciting} everywhere in $\inter(X)$, which guarantees that the process can escape in any direction. This yields part $(ii)$. When the network is bipartite, this property does not hold and we cannot guarantee that the process will not get stuck in an unstable set. 

Note that the bipartiteness condition does not imply that the process will not converge to an element of $Z^S$. However, we provide two examples in the appendix (Examples \ref{ex_1} and \ref{ex_2}) in which we show that the noise can vanish on bipartite networks in games that have either no strategic complements or no symmetric externalities. In our examples the noise vanishes at unstable equilibria.

\section{Locally ordinal potential games}\label{ssec:pot}

We introduce a class of games that we call the {\it locally ordinal potential games}. Recall that a game $\mathcal{G}$ is a \emph{potential game} ($PG$) if there is a function $P : X \rightarrow \mathbb{R}$ such that for all $x_{-i} \in X_{-i}$, for all $x_i, x'_i \in X_i$, we have
$u_i(x_i, x_{-i}) - u_i(x'_i, x_{-i}) = P(x_i, x_{-i}) - P(x'_i, x_{-i})$, and 
an \emph{ordinal potential game} ($OPG$) if $u_i(x_i, x_{-i}) - u_i(x'_i, x_{-i}) > 0 \iff P(x_i, x_{-i}) - P(x'_i, x_{-i}) > 0$.

\begin{definition}
A game $\mathcal{G}$ is a locally ordinal potential game ($LOPG$) if there is a differentiable function $P : X \rightarrow \mathbb{R}$ such that 
\begin{equation*}\label{eq:gopg}
\sgn \left(\frac{\partial u_i}{\partial x_i}(x)\right) = \sgn \left(\frac{\partial P}{\partial x_i}(x) \right)
\end{equation*}

\end{definition}

The class of $LOPG$ is large, in the sense that $PG \subset OPG \subset LOPG$ when $P$ is differentiable. It also contains many games of economic interest. For instance, both examples 1 and 2 are locally ordinal potential games. 

The generality of our results depends on the structure of the set of stationary points of the game under consideration, and in particular on whether it consists of isolated points or not. For instance, the public good game of example 1 generically has a finite number of isolated zeros, but can have continua of equilibria for certain values of the substitutability parameter\footnote{See \cite{BervFaure} for more details.}.

\begin{theorem}\label{th:pot_nash} Let $\mathcal{G}$ be an $LOPG$ and $P$ be sufficiently regular. Then
\begin{itemize} 
\item[$(i)$] $\mathbb{P} \left(\mathcal{L}(x_n)_n \subset Z\right) = 1.$
\item[$(ii)$] If $\mathcal{G}$ has isolated zeros, then
\[\mathbb{P} \left( \exists \; x^* \in NE: \; \, \lim_n x_n = x^* \right) = 1.\]
If, in addition, the interaction graph is non-bipartite on $NE$, then 
\[\mathbb{P} \left( \exists \; x^* \in NE^S: \; \, \lim_n x_n = x^* \right) = 1.\]
\end{itemize}
\end{theorem}

For any $LOPG$, the only set to which the stochastic learning process can converge is the set of zeros of $F$. Complex $\omega$-limit sets of the dynamical system, which are non-zeros, can be discarded. We cannot, however, be sure that the process will not reach a set containing other zeros, thus we cannot guarantee convergence to the set of Nash equilibria. When zeros are isolated, however, convergence to Nash is proved by the conjunction of the first point and the fact that the process cannot converge to an isolated other zero. 
Furthermore, we prove that the DGAP cannot converge to a linearly unstable Nash $\hat{x}$ if $\mathbf{G}(\hat{x})$ is non-bipartite (on this, we provide more details in Section \ref{ssec:isolated}).

When zeros are non-isolated, we cannot guarantee that the DGAP will converge to a stable set. We can use \cite{Ben99} to show that $\mathbb{P} \left(\mathcal{L}(x_n)_n \subset A  \right) >0$ on the event $\{x_0 \in \mathcal{B}(A)\}$, for any attractor  $A$ of the ODE \eqref{edo}, where $\mathcal{B}(A)$ is the basin of attraction of $A$. Combining this observation with point $(i)$ of Theorem \ref{th:pot_nash}, we get the following important implication: if a connected set $\Lambda$ is an attractor for $\dot{x} = F(x)$, then  $\Lambda$ is a connected component of $Z$.

However, when focusing on $LOPG$s, more can be said, since we are able to relate attractors of the dynamics to the potential function $P$, and to another dynamical system, extensively used in economics: Best-Response Dynamics (BRD). 

\begin{definition}
Let $\BR: X \rightarrow X, x \mapsto \BR(x) := (\BR_1(x_{-1}),...,\BR_n(x_{-n}))$. The \emph{continuous-time Best-Response dynamics} (thereafter, BRD) is defined as:  
\begin{equation} \label{eq:BRD}
\dot{x} = -x + \BR(x)
\end{equation}
\end{definition}

\begin{definition} Let $P$ be a smooth map and $\Lambda$ be a connected component of $Z$, we say that $\Lambda$ is a local maximum of $P$ if 
\begin{itemize}
\item[$(i)$] $P$ is constant on $\Lambda$: $P(x) = v, \; \forall x \in \Lambda$; 
\item[$(ii)$] there exists an open neighborhood $U$ of $\Lambda$ such that $P(y) \leq v \; \, \forall y \in U$
\end{itemize}
\end{definition}
\vspace{0.2cm}

\noindent We then have

\begin{theorem} \label{th:attractor} Assume $\mathcal{G}$ is an $LOPG$ and let $\Lambda$ be a connected set. Then the following statements are equivalent
\begin{itemize}
\item[$(i)$] $\Lambda$ is an attractor for $\dot{x} = F(x)$
\item[$(ii)$] $\Lambda$ is a local maximum of $P$
\item[$(iii)$] $\Lambda \subset NE$ and $\Lambda$ is an attractor for the best-response dynamics $\dot{x} = -x + \BR(x)$.
\end{itemize}
\end{theorem}

This result is positive and informative. First, it tells us that attractors are necessarily included in the set of Nash equilibria. Thus, although the process might converge to other zeros when stationary points are non-isolated, these points are unstable.

Second, Theorem \ref{th:attractor} provides two methods of finding the attractors:  one way is to look for local maxima of the potential function, which is very convenient when the function $P$ is known; and the other is to  look for attractors for another dynamics, possibly simpler to analyze, the BRD. 
Note that this second method establishes a relation between two dynamics that are conceptually unrelated. Indeed, the BRD assumes that agents are very sophisticated, as they know their exact payoff function, they observe their opponents' play and perform potentially complex computations. Solution curves may be very different, but surprisingly, both dynamics share the same set of attractors.

\section{Isolated zeros} \label{ssec:isolated}

In the two previous sections we did not assume any specific structure on the set of zeros of the dynamical system. However, in most economics games with continuous action spaces, the set of zeros, and in particular the set of Nash equilibria, would be finite. In that case, zeros are isolated points. For instance, in the public good game of example 1, \cite{BraKraDam14} show that the game has a finite number of equilibria for almost every value of substitutability between efforts. The same can be said about the games in example 2. In this section, we restrict our attention to these games. 

We start with a useful remark:

\begin{remark} \label{pr:isolated}  If $\hat{x} \in NE^{AS}$, then \[\mathbb{P} \left(\lim_n x_n =  \hat{x} \right) > 0\] on the event $\{x_0 \in \mathcal{B}(\hat{x})\}$.
\end{remark}

This is just a consequence of the result in \cite{Ben99} mentioned earlier and the fact that $\hat{x} \in NE^{AS}$ is an attractor. It says that the process can converge to desirable outcomes. We next turn to the hard part, i.e. excluding the convergence to undesirable zeros in every game with isolated zeros. 

\subsection{Non convergence to undesirable zeros}

In games with continuum of equilibria, we cannot exclude the possibility of our learning process getting arbitrarily close to elements of the set of other zeros. More precisely, there is no a priori reason to believe that the learning process will converge (to a point) when zeros of the dynamical system are connected components. If it does not, then  the process could come arbitrarily close to a continuum of $NE$ that is connected to a continuum of $OZ$, and oscillate between the two. However, when zeros are isolated this cannot happen and we can discard convergence to other zeros. Further, we can almost always 
discard convergence to linearly unstable Nash equilibria.

\begin{theorem} \label{th:isolated}  Let $\mathcal{G}$ be a game with isolated zeros and assume that $\hat{x} \in Z$. Then:
\begin{itemize}
\item[$i)$] If $\hat{x} \in OZ$, then $\mathbb{P} \left(\lim_n x_n =  \hat{x} \right) = 0.$ 
\item[$ii)$] If $\hat{x} \in \inter(X)$, $\hat{x} \in NE^{LU}$, and $\mathbf{G}(\hat{x})$ is non-bipartite, then
$\mathbb{P} \left(\lim_n x_n =  \hat{x} \right) = 0$
\end{itemize} 
\end{theorem}

The proof of point a) is a probabilistic proof. We show that in $OZ$, the players who are playing $0$ although they have a strictly positive gradient will, in expectation, increase their action level as they approach the boundary. This is of course a contradiction. 

\subsection{Concave games}\label{subs:conc_games}

As mentioned in Remark \ref{rk:rk1}, \cite{arrow1960stability} and \cite{rosen1965existence} analyzed similar dynamical systems in concave games. 
The first investigates a subclass of all games with payoff functions that are concave in players' own actions and convex in other players' actions. These games include the well-known class of zero-sum games. The authors then prove global convergence of  system \eqref{Syst_Arrow}. 

\cite{rosen1965existence} deals with concave games, and provides sufficient conditions for the game to have a unique Nash equilibrium when the strategy space is compact and convex: if there are some positive weights such that the weighted sum of the payoff functions is diagonally strictly concave, then the equilibrium of the game is unique. Under that assumption, the author proves that the weighted gradient system \eqref{Syst_Rosen} globally converges to this unique equilibrium. 

We are interested in determining whether the DGAP also converges in these games, but this raises several problems. First, we need to show that our deterministic system \eqref{edo} has the same good convergence properties as \eqref{Syst_Arrow} and \eqref{Syst_Rosen}. But this is not enough, since our process is stochastic, unlike theirs. Second therefore, we need to show that the limit set of the stochastic process \eqref{ode1} is included in the set of stationary points of the dynamical system \eqref{edo} for these games. Last, the games considered in \cite{arrow1960stability} sometimes have continua of equilibria. For instance, in zero-sum games, the set of equilibria is known to be convex. To avoid this issue, we maintain the concavity condition on the payoff functions but we require that at least one player's payoff function is strictly concave in own action. Under this assumption, we show that these games satisfy Rosen's (1965) condition - and thus have a unique Nash equilibrium. We next show that all games satisfying Rosen's condition have isolated zeros for our dynamical system. With this in hand, 
we prove that the DGAP converges to the unique Nash equilibrium with probability $1$.

Suppose that $u_i$ is concave in $x_i$ for every $i$. Following Rosen (1965), given $r \in (\mathbb{R}_+^*)^N$ and $x \in X$, let $g(x,r) \in \mathbb{R}^N$ be given by $g_i(x,r) = r_i \frac{\partial u_i}{\partial x_i}$\footnote{ The dynamical system $\dot{x} = g(x,r)$ is a weighted gradient system, and is significantly different from the system \eqref{edo}}. 
A game is \emph{diagonally strictly concave} if 
\begin{equation} \label{hy:rosen} 
 \exists r \in (\mathbb{R}_+^*)^N \, \mid \;  \forall x^0 \neq  x^1 \in X \, \text{ we have } \;  \left< x^1 - x^0 \mid g(x^0,r) \right> +  \left< x^0 - x^1 \mid g(x^1,r) \right> >0
 \end{equation} 
 Games having this property are denoted by $\mathcal{G}^{Ros}$.
It is proved (Theorem 2 of \cite{rosen1965existence}) that games in $\mathcal{G}^{Ros}$ have a unique Nash equilibrium when the state space is compact. In our context, where the state space is unbounded, they may have none.

Games considered by \cite{arrow1960stability} (which we call {\it concave-convex games}, and denote by $\mathcal{G}^{Arr}$) are as follow. Let $S$ be a subset of $N$, the set of players, and define $f^S = \sum_{i \in S} u_i - \sum_{i \in N\setminus S} u_i$. A game is concave-convex if $a)$ for each $S \subseteq N$, the function $f^S(x^S,x^{N\setminus S})$ is concave in $x^S$ for each $x^{N\setminus S}$ and convex in $x^{N\setminus S}$ for each $x^S$, and $b)$ for some $S^0 \subseteq N$, $f^{S^0}(x^{S^0},x^{N\setminus S^0})$ is strictly concave in $x^{S^0}$ for each $x^{N\setminus S^0}$. If in addition $u_i$ is strictly concave in $x_i$, then we say that the game is {\it strictly concave-convex}.

\begin{remark}\label{lm:arr_ros}
Stricly concave-convex games are diagonally strictly concave, i.e. $\mathcal{G}^{Arr} \subset \mathcal{G}^{Ros}$. Thus all properties of the later apply to the former.
\end{remark}

For simplicity, in the remainder of this section we will place ourselves in the setting of \cite{rosen1965existence}, i.e. we assume that the strategy space $X$ is a compact set. This guarantees that the Nash equilibrium is unique. When the set $X$ is unbounded, the game could have no equilibrium at all and if that happened, the process would go to infinity. Because this introduces unnecessary complexities in the proof, we restrict our attention to compact sets. 

The fact that the Nash equilibrium is unique is convenient for the study of dynamics where the Nash equilibria are the only stationary points. However, the system \eqref{edo} also has other zeros, since $Z = NE \cup OZ$. In the following theorem, we show that there is a finite number of other zeros, and thus all the stationary points are isolated. We also state our convergence result.

\begin{theorem}\label{th:rosen}
Let $\mathcal{G} \in \mathcal{G}^{Ros}$. Then,
\begin{itemize}
\item[$(i)$] $Z$ is a finite set
\item[$(ii)$] There is a  unique Nash equilibrium $\bar{x}$ and 
\[\mathbb{P}(\lim_n x_n = \overline{x}) = 1.\]
\end{itemize} 
\end{theorem}

The proof of the first point goes as follows:  we prove that games in $\mathcal{G}^{Ros}$ are such that, after removing a subset of players playing $0$, the remaining subgame is also in $\mathcal{G}^{Ros}$. Thus there is at most one Nash equilibrium for any combination of agents playing $0$. The number of such potential combinations is finite, so the result follows. 

In order to prove the second point of Theorem \ref{th:rosen}, we show that the zeros of \eqref{edo} are the only candidates for limit points of our process. We cannot do this in general games with isolated zeros, but in diagonally strictly concave games we can, by decomposing the state space into several subspaces (respectively, the interior of the space and every face) and constructing appropriate Lyapunov functions for each subspace. As a consequence, we prove that every solution of \eqref{edo} converges to one of the zeros. Since zeros are the only candidates, we get the desired conclusion by using point i) of Theorem \ref{th:isolated}. 

\appendix 

\addcontentsline{toc}{section}{Appendices}
\section*{Appendix}

\section{Proof of results of Section \ref{model}}

\paragraph{\underline{Proof of Proposition \ref{pr:stochapp}.}} We first prove that the process can be written as in equation \eqref{ode1}. Second we prove that the process is well-defined, i.e. $x_n^i > 0$ for all $i$ and all $n$. \\

1- We have, for any $i \in \mathcal{N}$,
\[e_{2n+2}^i - e^i_{2n} = e^i_{2n} \epsilon_n^i \Delta u^i_{n+1}\]

A first order development gives
\begin{eqnarray*}
\epsilon_n^i \Delta u^i_{n+1} &=& \epsilon_n^i \left(u_i \left(e^i_{2n} + \frac{1}{n+1} \epsilon_n^i ,e^{-i}_{2n}+ \frac{1}{n+1} \epsilon_n^{-i} \right) - u_i(e^i_{2n},e^{-i}_{2n}) \right)\\
&=& \frac{1}{n+1} (\epsilon_n^i)^2 \frac{\partial u_i}{\partial x^i}(e_{2n}) + \frac{1}{n+1} \epsilon_n^i \sum_{j \neq i}\epsilon_n^j  \frac{\partial u_i}{\partial x^j}(e_{2n}) + \mathcal{O}(\frac{1}{n^2})
\end{eqnarray*}
Because $(\epsilon_n^i)^2 = 1$ and $x_n = e_{2n}$, we have
\begin{eqnarray*}
x_{n+1}^i - x_n^i &=& \frac{1}{n+1} x_n^i \frac{ \partial u_i}{\partial x^i}(x_n) + \frac{1}{n+1} \epsilon_n^i x_n^i \sum_{j \neq i}\epsilon_n^j \frac{\partial u_i}{\partial x^j}(x_n) + \mathcal{O}(\frac{1}{n^2})
\end{eqnarray*}
By setting $U_{n+1}^i = \epsilon_n^i x_n^i \sum_{j \neq i}\epsilon_n^j   \frac{\partial u_i}{\partial x^j}(x_n)$, we get equation \eqref{ode1}.  Finally, note that $\mathbb{E} \left(\epsilon_n^j\right) = 0$ for all $j$, and that $\epsilon_n^i$ and $\epsilon_n^j$ are independent, so that
\[\mathbb{E} \left(U_{n+1} \mid \mathcal{F}_n \right) = 0. \;\;\blacksquare\]

2- Let us now show that the process is well-defined. 
Notice that Hypothesis~\ref{hyp:conc} implies that $Du_i$ is bounded everywhere. For simplicity and without loss of generality, we will assume that $|u_i(x)-u_i(x')| < \|x-x'\|_{\infty}$. This is just for simplicity, the proof can easily be accommodated otherwise.
Let $n \geq 0$. By assumption, $|u_i(x) - u_i(x')| \leq \|x-x'\|_{\infty}$ for all $i$. Thus, 
\[\frac{x_{n+1}^i}{x_n^i} \geq (1 - \|e_{2n+1} - x_n\|_{\infty}),\]
and $|e_{2n+1}^i - x_n^i| \leq \frac{1}{n+1}$ for all $i$. As a consequence,
\[\frac{x_{n+1}^i}{x_n^i} \geq (1 - \frac{1}{n+1}).\]
Thus, $x_1^i \geq 0$
and
\[x_n^i \geq x_1^i \prod_{k=1}^{n-1} \left(1 - \frac{1}{k+1}\right) = \frac{1}{n+1} x_1^i \geq 0 .\]
\\
Note that at the beginning of the process, steps are large. Thus in case $x^i_0$ is close to $0$, the exploration phase might take players to the negative
orthant ($e_{1} \leq 0$). This can only happen because the first steps are large. In order to avoid that, we can either assume that $x_0^i > 1$ (i.e. players start far enough from the boundary), or that the process begins at step $n \geq \min_i \{E(1/x_0^i)\} + 1$, where
$E(a)$ is the integer part of $a$ (i.e. the first steps are not too large). In any case, this is totally innocuous for
what we do and guarantees that $e_{1} \geq 0$. $\; \; \blacksquare$
\vspace{.3cm}

\paragraph{\underline{Proof of Proposition~\ref{pr:OZ_unst}}.} Pick $\hat{x} \in OZ$ and assume without loss of generality that $\hat{x}_1 = 0$ with $\frac{\partial u_1}{\partial x_1}(\hat{x}) >0$. Then 
\[\frac{\partial F_1}{\partial x_1}(\hat{x}) = \frac{\partial u_1}{\partial x_1}(\hat{x}), \; \, \mbox{ and } \; \, \frac{\partial F_1}{\partial x_j}(\hat{x}) = 0 \; \text{ for } \; j \neq 1.\]
Hence $\frac{\partial u_1}{\partial x_1}(\hat{x}) \in \Sp(DF(\hat{x}))$, and the associated eigenvector is $v = (1, 0,...,0)$ which points inwards (i.e. $\hat{x} + v \in X$). Thus, necessarily $OZ \subset Z^{LU}$. 

Next, $Z^S \subset NE$ is a consequence of $Z = NE \cup OZ$, $Z^S = Z \setminus Z^{LU}$ and $OZ \subset Z^{LU}$.
 $\; \; \blacksquare$
\vspace{.2cm}

\section{Proof of results of Section \ref{ssec:complements}} \label{appendix:complements}

\subsection{Proof of Proposition \ref{lm:nonconvboundary}}

\noindent Under Assumption~\ref{hy:boundary}, for any $i$, there exists $\overline{x}_i >0$ such that 
\[\frac{\partial u_i}{\partial x_i}(x_i,0) > \alpha_i >0, \; \forall x_i \leq \overline{x}_i.\]
Since the game has strategic complements,
\begin{equation}\label{aux_1}
\frac{\partial u_i}{\partial x_i}(x_i,x_{-i}) > \alpha_i >0, \; \forall x_i < \overline{x}_i, \; \forall x_{-i} \in X_{-i}.
\end{equation}
As a consequence, any solution trajectory with initial condition in the set $\{x \in X: x_i \in ]0,\overline{x}_i[\}$ is in the set $\{x \in X: x_i > \overline{x}_i\}$ after some finite time $t >0$. Let $a = \min_i \overline{x}_i$. Therefore any invariant set is contained either in $ [a,\infty[^{N}$ or in $\partial X$. Thus, by the aforementioned result of Bena\"{i}m (1999), we can conclude that 
 $$\mathbb P \left (\mathcal{L}((x_n)_n) \subset [a,\infty[^{N} \right )+ \mathbb P \left (\mathcal{L}((x_n)_n) \subset \partial X \right )=1.$$ 
 
 In what follows we will show that $\mathbb P \left (\mathcal{L}((x_n)_n) \subset \partial X \right )=0$. The main idea is to exploit the fact that the strategic complementarity condition implies that, if $x \in \partial X$ and for some coordinate $x_i=0$ then $\frac{\partial u_i}{\partial x^i}(x)$ must be strictly positive (there is no Nash equilibria on $\partial X$).  
  \begin{remark}\label{rem:app_preuve}
 Three simple observations are in order.  \begin{itemize}
 \item[$(i)$] Condition \eqref{aux_1} implies that if $\frac{\partial u_i}{\partial x^i}(x)\leq 0$, then $x^i\geq a$.
  \item[$(ii)$] If $x \in  X \setminus  [a,+\infty[^N$, the set of coordinates for which $\frac{\partial u_i}{\partial x^i}(x)>0$, $I_x$, is always nonempty. This is because if for all $i \in \{ 1,\ldots, N\}$,  $\frac{\partial u_i}{\partial x^i}(x)\leq 0$ then $x \in [a,+\infty[^N$.
  \item[$(iii)$]Moreover, also from \eqref{aux_1}, the coordinate $k$ achieving the minimum value of a vector $x \in X \setminus  [a,+\infty[^N$ verifies that $\frac{\partial u_{k}}{\partial x^{k}}(x)>\alpha$, where $\alpha= \min_i \alpha_i>0$. Therefore this particular $k$ belongs to the set  $I_x$.
  \end{itemize}
\end{remark}
  
Let $d(x,\partial X)$ be distance for the infinity norm of $x$ to $\partial X$, i.e. $d(x,\partial X)=\min_i x_i$.
 Let us take  $R>a$ and consider the sets  $\mathcal U_R$: 
 \begin{equation*}
   \text{ and }\mathcal U_R = \left \{x \in X \,; \,\frac{\partial u_i}{\partial x^i}(x) < 0 \Rightarrow -x_i \frac{\partial u_i}{\partial x^i}(x) \leq R  \right \}.
 \end{equation*}
 Observe that $\partial X$ can be written as an increasing union of the form:
  \begin{equation}\label{aux_eps}
 \partial X = \bigcup_{R=1}^{\infty} \left(\partial X\cap \mathcal U_{R}\right),
 \end{equation}
In order to show that $\mathbb P \left (\mathcal{L}((x_n)_n) \subset \partial X \right )=0$, it is sufficient to prove that, for all $R>a$, 
$\mathbb P \left (\mathcal{L}((x_n)_n) \subset \partial X \cap \mathcal U_{R} \right )=0$. By contradiction, assume that there exists $R>a$ such that  $\mathbb P \left (\mathcal{L}((x_n)_n) \subset \partial X \cap \mathcal U_{R} \right )>0$ and let  $0<\varepsilon<a$. On the event $\{ \mathcal{L}((x_n)_n) \subset \partial X \cap \mathcal U_{R}\}$, there exists a (random) $n_* \in \mathbb N$ such that
\begin{equation}\label{event_aux}
 \mathbb P \left (\left \{\mathcal{L}((x_n)_n) \subset\partial X \right \}  \cap \left \{ x_n \in V_{\varepsilon}\cap \mathcal U_{R}, \text{ for all } n \geq n_* \right \} \right )>0,
\end{equation}
where
 \[\mathcal V_\varepsilon= \left \{x \in X \,; \,d(x,\partial X) \leq\varepsilon \right \}.
 \]
 
In what follows, we work on the event $E$ defined by \eqref{event_aux} and we assume that $n \geq n_*$.

 For $\beta >0$, let the function 
 \[
 \Phi_\beta (x)=- \frac{1}{\beta}\ln \left ( \sum_{i=1}^N \exp(-\beta x^i)\right ),
 \]
 which is concave if extended as $-\infty$ to $\mathbb R^N$. The function $\Phi$ verifies the well-known relation
  \begin{equation}\label{eq:relation_app}
\min_{i=1,\ldots,N}x^i -\frac{\ln(N)}{\beta}\leq \Phi_\beta (x)\leq \min_{i=1,\ldots,N}x^i.
  \end{equation}

 From a straightforward calculation we have that, for all $i \in \{1,\ldots, N\}$,
 \[
 \frac{\partial \Phi_\beta}{\partial x^i}(x)= \pi_i(x) , \text{ where }  \pi_i(x)=\frac{\exp(-\beta x^i)}{\sum_{j=1}^N\exp(-\beta x^j)}. 
  \]
 Also, for all $i,j \in \{1,\ldots, N\}$
 \[
 \frac{\partial^2 \Phi_\beta}{\partial x^j \partial x^i}(x)= -\beta\pi_i(x)(\delta_{ij} - \pi_j(x)),
 \]
where $\delta_{ij}=1$ if  $i=j$, and $0$ otherwise. This implies that $\nabla \Phi_\beta$ is $L$-Lipschitz. In fact, $L\leq 2\beta$ for the infinity norm.

 Observe that if $x \in V_{\varepsilon}\cap \mathcal U_{R}$ and if $\frac{\partial u_i}{\partial x^i}(x) \leq 0$ for some coordinate $i$, we have that 
 \begin{equation*}
 \pi_i(x)\leq \exp \left (-\beta( a -\varepsilon )\right ),
 \end{equation*}
 using the fact that it exists some $k$ such that $x^k \leq \varepsilon$ and that  $x^i \geq a$ ({\em c.f.} Remark \ref{rem:app_preuve}). 
 
 On the other hand, for $k \in I_x$ such that $x^k=\min_i x^i$,
 \begin{equation*}
 \pi_k(x)= \frac{1}{1+\sum_{j\neq k}\exp \left ( \beta x^k -\beta x^j \right ) }\geq  \frac{1}{N}.
 \end{equation*}

Recall that the variable $x_n$ follows the recursion
\[x_{n+1}^i = x_n^i  + \frac{1}{n+1} \left( x_n^i\frac{\partial u_i}{\partial x_i}(x_n) + U_{n+1}^i + \xi^i_{n+1} \right),\]
where $\mathbb{E} \left(U_{n+1} \mid \mathcal{F}_n \right) = 0$ and $|\xi^i_{n}| \leq C/n$, for a deterministic constant $C$.

 Let us define $z_n=\Phi_\beta(x_n)$. Note first that, from equation \eqref{eq:relation_app},
\begin{equation}\label{aux_zn}
- \frac{\ln(N)}{\beta} \leq \min_{i=1,\ldots,N}x_n^i -  \frac{\ln(N)}{\beta}\leq z_n\leq \min_{i=1,\ldots,N}x_n^i \leq \epsilon.
\end{equation}
Consequently, $\mathcal L ((z_n)_n) \subset [-\ln(N)/\beta,0]$ almost surely on $E$.

On the other hand, since the function $-\Phi_\beta$ is convex with $L$-Lipschitz gradient, we have that
\[
-\Phi_\beta(x_{n+1}) \leq -\Phi_\beta(x_{n}) +\langle -\nabla\Phi_\beta(x_{n})\,,\, x_{n+1} - x_n \rangle + \frac{L}{2}\norm{x_{n+1} -x_n}^2.
\]
Equivalently,
 \begin{equation*}
  \begin{aligned}
z_{n+1} &\geq z_n + \sum_{j=1}^N \pi_j(x_n)(x_{n+1}^j - x_n^j)-  \frac{L}{2}\norm{x_{n+1} -x_n}^2,\\
&=z_n +\frac{1}{n+1}\sum_{j=1}^N\pi_j(x_n) \left( x_n^j\frac{\partial u_i}{\partial x_j}(x_n) + U_{n+1}^j + \xi^j_{n+1} \right) -  \frac{L}{2}\norm{x_{n+1} -x_n}^2,\\
&\geq z_n +\frac{1}{n+1}\sum_{j=1}^N\pi_j(x_n) x_n^j\frac{\partial u_i}{\partial x_j}(x_n)+ \frac{1}{n+1}\sum_{j=1}^N\pi_j(x_n) U_{n+1}^j - \frac{c}{(n+1)^2},
  \end{aligned}
 \end{equation*}
 for some deterministic constant $c\geq 0$.
Therefore, taking conditional expectation and omitting the quadratic term,
 \begin{equation*}
\mathbb{E} \left ( z_{n+1} \mid \mathcal{F}_n \right) \geq  z_n+\frac{1}{n+1}\sum_{j=1}^N\pi_j(x_n) x_n^j\frac{\partial u_i}{\partial x_j}(x_n).
 \end{equation*}
 Recall that $I_{x_n}$ is the set of indices such that $\frac{\partial u_i}{\partial x^i}(x_n)>0$ and that $k_n$ the coordinate giving the minimum of $x_n$ is in $I_{x_n}$ and verifies moreover that $\frac{\partial u_i}{\partial x^i}(x_n)>\alpha$ . Let $J_{x_n}$ the set of indices such that $\frac{\partial u_i}{\partial x^i}(x_n)\leq 0$. 
  
 For all $n \geq n_*$ we have
  \begin{equation*}
  \begin{aligned}
\mathbb{E} \left ( z_{n+1} \mid \mathcal{F}_n \right) &\geq  z_n+\frac{1}{n+1}\sum_{j \in I_{x_n}}\pi_j(x_n) x_n^j\frac{\partial u_i}{\partial x_j}(x_n) +\frac{1}{n+1} \sum_{j \in J_{x_n}}\pi_j(x_n) x_n^j\frac{\partial u_i}{\partial x_j}(x_n),\\
&\geq  z_n+\frac{z_n}{n+1}\frac{\alpha}{N}+ \frac{1}{n+1}\sum_{j \in J_{x_n}}\pi_j(x_n) x_n^j\frac{\partial u_i}{\partial x_j}(x_n),
  \end{aligned}
 \end{equation*}
 using that $x_n^{k_n}\geq z_n$ and that $\pi_{k_n}(x_n)\geq 1/N$. On the other hand, using the definition of $\mathcal U_R$, we obtain
 \[
\sum_{j \in J_{x_n}}\pi_j(x_n) x_n^j\frac{\partial u_i}{\partial x_j}(x_n)  \geq -|J_{x_n}| R\exp \left (-\beta( a -\varepsilon )\right )\geq - NR \exp \left (-\beta( a -\varepsilon )\right ).
 \]
Thus 
   \begin{equation*}
  \begin{aligned}
 \mathbb{E} \left ( z_{n+1} \mid \mathcal{F}_n \right) &\geq  z_n+\frac{1}{n+1}\left (\frac{\alpha}{N}z_n -  NR\exp \left (-\beta( a -\varepsilon )\right ) \right ).\\
  \end{aligned}
 \end{equation*}
Let us consider the change of variables 
 $$\theta_n= \left (z_n + \frac{\ln(N)}{\beta} \right ) \geq \min_{i}x_n^i \geq 0.$$
 Then, 
 \begin{equation*}
 \mathbb{E}( \theta_{n+1} ) \geq   \mathbb{E}(\theta_n)+ \frac{\alpha}{N}\frac{1}{n+1}\left ( \mathbb{E}(\theta_n)  -  \underbrace{\left \{\frac{N^2}{\alpha}R\exp\left (-\beta( a -\varepsilon )\right )  +  \frac{\ln(N)}{\beta} \right \}}_{c(\beta)} \right ).
 \end{equation*}
 
 Let us note that $\mathbb E(\theta_{n_*})>0$ since $\min_{i}x_{n_*}^i \geq 1/(n_*+1)$ almost surely. Now, we can fix $\beta>0$ sufficiently large such that $0< c(\beta)< \mathbb E(\theta_{n_*})$. So that
 \[
 \mathbb{E}( \theta_{n+1}) \geq   \mathbb{E}(\theta_n)+  \frac{\alpha}{N}\frac{1}{n+1}\left (\mathbb{E}(\theta_n) - c(\beta)\right ).
 \]
 Let us call $\rho_n=\mathbb{E}(\theta_n) - c(\beta)$. Then, we want to analyse the recursion $\rho_{n+1}\geq \rho_n(1 +  \frac{\alpha}{N}\frac{1}{n+1})$, with $\rho_{n_*} >0$. Hence, for $n \geq n_*$,
 \[
 \rho_{n+1}\geq \rho_{n_*} \prod_{i=n_*}^{n}\left (1 +  \frac{\alpha}{N}\frac{1}{i+1} \right),
 \]
 where the right-hand-side goes to infinity. Finally, we can conclude that $\mathbb E(z_n)$ goes to infinity, which is a contradiction with the fact that $z_n \in [-\ln(N)/\beta,\epsilon]$ almost surely on the event $E$.
 $\; \; \blacksquare$
\vspace{.3cm}

\subsection{Proof of Theorem \ref{th:convcompl}}
The proof of Theorem \ref{th:convcompl} involves several arguments. For point $(i)$, we use a result from \cite{Pem90}, while for point $(ii)$ we adapt a result from  \cite{BenFau12}. For both points we use Proposition \ref{lm:nonconvboundary}, i.e the fact that the limit set of the process cannot include points on the boundary of the state space.

\noindent For the first point of Theorem \ref{th:convcompl}, let us recall some results on non convergence. Let $\tilde{x}$ be a linearly unstable equilibrium. Assume without loss of generality that the unstable space at $\tilde{x}$ is one-dimensional, that is $DF(\tilde{x})$ has only one  eigenvalue $\mu$ with positive real part, and we call $v$ the associated normalized eigenvector. We use a result from \cite{Pem90}, more precisely in the settings of \cite{BraDuf96}, which states that a sufficient condition for non convergence to $\tilde{x}$ is that the noise is exciting in the unstable direction, i.e.:
\begin{equation} \label{eq:noise}
\liminf_{n \rightarrow + \infty} \mathbb{E} \left(\langle U_{n+1},v \rangle^2 \mid  \mathcal{F}_n \right) > 0
\end{equation}
on the event $\{\lim_n x_n = \tilde{x} \}$. 

Consider any $x_n$ and any vector $v$. Then
\begin{equation*}
\langle U_{n+1},v \rangle^2 = \left(\sum_{i<j} \epsilon_n^i \epsilon_n^j  \left(v_i x_n^i  \frac{\partial u_i}{\partial x^j}(x_n) + v_j x_n^j \frac{\partial u_j}{\partial x^i}(x_n)\right) \right)^2 
\end{equation*}
Using $\mathbb{E} \left( \epsilon_n^i \epsilon_n^j \right) = 0$ if $i \neq j$ and $(\epsilon_n^i)^2  = 1$, we get 
\begin{eqnarray}\label{eq:noise_unif}
\mathbb{E} \left(\langle U_{n+1},v \rangle^2 \mid  x_n \right) 
&=& \sum_{i<j}  \left(v_i x_n^i \frac{\partial u_i}{\partial x^j}(x_n) + v_j x_n^j \frac{\partial u_j}{\partial x^i}(x_n)\right)^2 
\end{eqnarray}
\vspace{.2cm}

\paragraph{\underline{Proof of Theorem \ref{th:convcompl} (i)}.} Let $\tilde{x}$ be a linearly unstable equilibrium. We want to show that 
\[\sum_{i<j} \left(v_i \tilde{x}^i \frac{\partial u_i}{\partial x^j}(\tilde{x}) + v_j \tilde{x}^j \frac{\partial u_j}{\partial x^i}(\tilde{x})\right)^2 \neq 0.\]
where $v$ is the normalized eigenvector associated associated to the unstable direction of $\tilde{x}$. 

Note that $\tilde{x} \notin \partial X$ by Proposition \ref{lm:nonconvboundary}, and that when $\tilde{x} \in \inter(X)$ and the interaction graph is non-bipartite, then the result is a direct implication of 
Theorem \ref{th:isolated} in Section \ref{ssec:isolated}. Thus, here we assume that the interaction graph is connected and bipartite, and that $\tilde{x} \in \inter(X)$. This implies that there exists a partition $(A,B)$ of $N$ such that if $a \in A$ and $\frac{\partial u_a}{\partial x_b}(\tilde{x})\frac{\partial u_b}{\partial x_a}(\tilde{x})>0$ then $b \in B$.

Using the computations just developed, we need to show that 
\begin{equation}\label{eq:bip_comp}
\sum_{a<b}  \left(v_a x^a \frac{\partial u_a}{\partial x^b}(\tilde{x}) + v_b x^b  \frac{\partial u_b}{\partial x^a}(\tilde{x})\right)^2 \neq 0.
\end{equation}
Assume the contrary. Then we must have $v_a x^a \frac{\partial u_a}{\partial x^b}(\tilde{x}) + v_b x^b  \frac{\partial u_b}{\partial x^a}(\tilde{x}) = 0$ for all $a \in A$ and all $b \in B$. Because $x^i > 0$ for all $i$, and by Hypothesis \ref{hyp:sym_ext} (symmetric externalities), it must be that 
$\sgn(v_a) = - \sgn(v_b)$ for any $a \in A$ and any $b \in B$. Since the interaction graph is connected, we may assume without loss of generality that $v_a>0 \; \forall \, a \in A$ and $v_b < 0 \;  \forall \, b \in B$. 

Because $\mu$ is strictly positive and $v$ is the corresponding normalized eigenvector, we should have $\langle v DF(\tilde{x}), v \rangle = \mu \sum_i v_i^2 > 0$, since $v \neq 0$.
However, we will show that this can only be true if equation \eqref{eq:bip_comp} holds.  
By a simple rearrangement of the indexes, the Jacobian matrix at $\tilde{x}$ can be written as follows: 
\[DF(\tilde{x}) = \begin{pmatrix}
D_A & M \\
N & D_B 
\end{pmatrix},\]
where $D_A$ is diagonal and the diagonal terms are equal to $x^a \partial^2 u_a/\partial (x^a)^2(\tilde{x}) \leq 0$ with $a \in A$; and similarly for $D_B$. $M$ and $N$ are non-negative matrices, as $x^i \partial^2 u_i/ \partial x^i \partial x^j \geq 0 \; \forall i \neq j$. 

Thus,
\[\left<v DF(\tilde{x}), v \right> = \sum_i v_i^2 x^i \partial^2 u_i/ \partial (x^i)^2 + \sum_{a \in A,b \in B} v_a v_b \left(x^a \frac{\partial^2 u_a}{\partial x^a \partial x^b} + x^b \frac{\partial^2 u_b}{\partial x^a \partial x^b}\right) \leq 0,\]
a contradiction. To see why this inequality holds, remember that the terms in the first sum are all negative by Hypothesis \ref{hyp:conc} and the fact that $\tilde{x}$ is a Nash equilibrium. The terms in the second sum are also all negative since $v_a.v_b < 0$ and by strategic complements.
 $\; \; \blacksquare$\\
\vspace{.2cm}

For the proof of point ($ii$), we use the following theorem of \cite{BenFau12},  conveniently adapted to our setting.

\begin{theorem}[Bena\"im and Faure, 2012]\label{th:hyperb} Let $(x_n)_n \in X$ be a random process that can be written as
\begin{equation*}
x_{n+1} = x_{n} +\frac{1}{n+1} \left(F(x_{n}) + U_{n+1}+ \xi_{n+1}  \right)
\end{equation*}
where
\begin{itemize}
\item[$(i)$]  $F: X \rightarrow \mathbb R^N$ is a smooth map, that is cooperative and irreducible in $\inter(X)$,
\item[$(ii)$] $U_{n+1}$ is a bounded martingale difference and is uniformly exciting, {\em i.e.} the matrix
\[\mathbb{E} \left(U_{n+1} U^T_{n+1} \mid x_n = x \right)\]
is positive definite for any $x \in \inter(X)$,
\item[$(iii)$] $\xi_n = \mathcal{O}(1/n)$, and
\item[$(iv)$] there exists $a >0$ such that $\mathcal{L}(x_n)_n \subset [a,+ \infty[$ almost surely.
\end{itemize}
Then 
\[\mathbb{P} \left(\exists \, x^* \in Z^S: \; \, \lim_n x_n = x^* \right) = 1\]
on the event $\{\limsup_n \|x_n\| < + \infty \}$.
\end{theorem}
\vspace{.2cm}

\paragraph{\underline{Proof of Theorem \ref{th:convcompl} (ii)}.} We want to apply Theorem~\ref{th:hyperb}. 

When the game is of strategic complements, our dynamics $\dot{x} = F(x)$ is \emph{cooperative} because all non-diagonal entries of $DF(x)$ are nonnegative. In addition, Hypothesis \ref{hyp:sym_ext} guarantees that the interaction graph is strongly connected. Thus the matrix $DF(x)$ is {\em irreducible} for any $x$ in the interior of $X$. These two facts provide point $(i)$.  
Points $(iii)$ and $(iv)$  follow from Propositions \ref{pr:stochapp} and \ref{lm:nonconvboundary}, respectively.
To prove point $(ii)$, we prove that if a network is non-bipartite and the game exhibits symmetric interactions, then the noise is uniformly exciting. 

Since for any $v \in \mathbb R^N$ we have that $v^TU_{n+1} U_{n+1}^Tv=\langle U_{n+1},v \rangle^2 \geq 0$, $(ii)$ amounts to showing that, for any $x \in \inter(X)$, we have 
\[\mathbb{E} \left(\langle U_{n+1},v\rangle^2 \mid  x_n = x \right)= 0\] 
if and only if $v=0$. By equation \eqref{eq:noise_unif}, we see that the condition is verified in $v$ if and only if 
\[\forall i < j, \; \, v_i \tilde{x}^i \frac{\partial u_i}{\partial x^j}(\tilde{x}) + v_j \tilde{x}^j \frac{\partial u_j}{\partial x^i}(\tilde{x}) = 0.\]
We now prove that under the assumption of symmetric interactions and non-bipartiteness of the graph, this quantity is positive.

Since the interaction graph is non bipartite in $\tilde{x}$, there is at least one odd cycle. Let us assume, for simplicity but without loss of generality, that this cycle is of length $3$: there exist $i,j,k$ such that
\[\frac{\partial u_i}{\partial x_j}(\tilde{x}) \frac{\partial u_j}{\partial x_i}(\tilde{x}) >0, \frac{\partial u_j}{\partial x_k}(\tilde{x}) \frac{\partial u_k}{\partial x_j}(\tilde{x}) >0, \frac{\partial u_i}{\partial x_k}(\tilde{x}) \frac{\partial u_k}{\partial x_i}(\tilde{x}) >0.\]
We thus have 
$\sgn(v_i) = - \sgn(v_j) = \sgn(v_k) = - \sgn(v_i)$ which implies, since $\tilde{x}$ is interior, that $v_i=v_j=v_k=0$. As a consequence, for every agent $l$ linked to $i,j$ or $k$, we must have $v_l = 0$. Recursively, since the interaction graph is connected, we must have $v=0$, which concludes the proof. $\; \; \blacksquare$  
\vspace{0.2cm}

\section{Proof of results of Section \ref{ssec:pot}.}

\subsection{Proof of Theorem~\ref{th:pot_nash}}

Before  proving  Theorem \ref{th:pot_nash}, let us define the following dynamical concept: 
\begin{definition} Let $P:X \rightarrow \mathbb{R}$ be continuously differentiable. We say that $P$ is a strict\footnote{Generally, $P$ is a Lyapunov function for $\dot{x} = F(x)$ with respect to $\Lambda$ if $t \mapsto P(\varphi(x,t))$ is constant on $\Lambda$ and strictly increasing for $x \notin \Lambda$; when the component $ \Lambda$ coincides with the set of stationary points of the flow, then we say that $P$ is strict.} Lyapunov function for $\dot{x} = F(x)$ if 
\begin{itemize}
\item for $x \in Z$ the map $t \mapsto P(\varphi(x,t))$ is constant;
\item for $x \notin Z$ the map $t \mapsto P(\varphi(x,t))$ is strictly increasing.
\end{itemize} 
\end{definition}

\begin{lemma} \label{lm:gopg} Assume that $\mathcal{G}$ is an $LOPG$  with continuously differentiable potential $P$. Then 
\begin{itemize}
\item[$(i)$] $P$ is a strict Lyapunov function for $\dot{x} = -x + \BR(x)$.
\item[$(ii)$] $P$ is a strict Lyapunov function for $\dot{x} = F(x)$ (where $F_i(x)=x_i \frac{\partial u_i}{\partial x_i}(x)$).
\end{itemize}
\end{lemma}

\hop {\bfseries Proof.} By assumption,
\[\forall x, \, \forall i, \; \, \frac{\partial u_i}{\partial x_i}(x)>0 \Rightarrow \frac{\partial P}{\partial x_i}(x) > 0  \mbox{ and } \; \, \frac{\partial u_i}{\partial x_i}(x) < 0 \Rightarrow \frac{\partial P}{\partial x_i}(x) < 0.\]
\noindent $(i)$ We have
\[\left<DP(x),-x+\BR(x) \right> = \sum_i \frac{\partial P}{\partial x_i}(x) (-x_i + \BR_i(x)).\]
We need to check that, if $x \notin NE$, then this quantity is positive. Let $i$ be such that $x_i \neq \BR_i(x)$, say $x_i < \BR_i(x_{-i})$. Then by strict concavity of $u_i$ we have $\frac{\partial u_i}{\partial x_i}(x) > 0$. Thus $\frac{\partial P}{\partial x_i}(x) >0$ and $\left<DP(x),-x+\BR(x) \right>>0.$
\vspace{.2cm}

\noindent $(ii)$ We have
\[ \left<DP(x),F(x) \right> = \sum_i x_i \frac{\partial u_i}{\partial x_i}(x) \frac{\partial P}{\partial x_i}(x) .\]
We need to check that, if $x \notin Z$, then this quantity is positive. Let $i$ be such that $F_i(x) \neq 0$. Then  $x_i >0$ and  $\frac{\partial u_i}{\partial x_i}(x) \neq 0$, which implies that
\[x_i \frac{\partial u_i}{\partial x_i}(x) \frac{\partial P}{\partial x_i}(x) >0.\]  
and the proof is complete. $\; \; \blacksquare$

\vspace{1cm}

\paragraph{\underline{Proof of Theorem \ref{th:pot_nash} (i)}.} For this part, we use the general result given by Proposition 6.4 in \cite{Ben99}, which asserts that if $P$ is a strict Lyapunov function with respect to $Z$ and $P(Z)$ has empty interior, then $\mathcal L(x_n)\subset Z$ almost surely.  
Therefore, the following lemma finishes the proof. $\; \; \blacksquare$

\begin{lemma} \label{lm:emptyint} Assume $\mathcal{G}$ is an $LOPG$. Then, if $P$ is sufficiently regular, $P(Z)$ has an empty interior.
\end{lemma}
\hop {\bfseries Proof.} We decompose the set of zeroes of $F$ as a finite union of sets on which we can use Sard's Theorem.
\vspace{.2cm}

\noindent Let $A$ be any subset of agents and $Z_A$ be the set 
\[\left\{x \in Z: \; x_i=0 \; \, \forall i \notin A, \; \; \frac{\partial u_i}{\partial x_i}=0 \; \forall i \in A \right\}.\]
It is not hard to see that $Z_A$ is closed. Moreover $Z = \cup_{A \in \mathcal{P}(\{1,...,N\})} Z_A$.

\noindent We now prove that $P$ is constant on $Z_A$. Let $P^A:[0,1]^A \rightarrow \mathbb{R}$ be defined as
\[P^A(z) := P(z,0).\]
For $x \in Z_A$, denote by $x^A=(x_i)_{i \in A}$. We then have $P^A(x^A) = P(x)$. Moreover, for $i \in A$, 
\[\frac{\partial P^A}{\partial x_i} = 0\]
by definition of $Z_A$ and the additional assumption we made on $P$. Hence  
\[\{x^A: \; \, x \in Z_A\} \subset \{z \in [0,1]^A: \; \, \nabla_z P^A = 0\}.\] 
Now $P$ is sufficiently differenciable, so is $P^A$, and by Sard's Theorem, $P^A(\{x^A: \; \, x \in Z_A\})$ has empty interior in  $\mathbb{R}^A$. As an immediate consequence, $P^A$ is constant on $\{x^A: \; \, x \in Z_A\}$ , which directly implies that $P(Z_A)$ has empty interior. Since $Z$ is a finite union of such sets,  $P(Z)$  has empty interior. $\; \; \blacksquare$
\vspace{.2cm}

\paragraph{\underline{Proof of Theorem \ref{th:pot_nash} (ii)}.} This proof relies on Theorem \ref{th:isolated} in Section \ref{ssec:isolated}, and on the following: by Lemma \ref{lm:gopg} and  Corollary 6.6 in \cite{Ben99}, we have
\[ \mathbb{P} \left(\exists \hat{x} \in Z \,\text{ such that } \; \lim_n x_n = \hat{x}\right) = 1. \]
Because convergence to the zeroes is guaranteed, Theorem \ref{th:isolated} gives us the result. Point a) of Theorem \ref{th:isolated} gives us the first point of 
Theorem \ref{th:pot_nash} (ii), and the second point of 
Theorem \ref{th:pot_nash} (ii) is a consequence of point b) of Theorem \ref{th:isolated}. $\; \; \blacksquare$

\vspace{.3cm}

\subsection{Proof of Theorem~\ref{th:attractor}}

 First we prove that $(i)$ implies $(ii)$.  Since $\mathcal{G}$ is an LOPG, $P(Z)$ has empty interior (see Lemma~\ref{lm:emptyint} above). Moreover, we have $\Lambda \subset Z$. Thus $P$ is constant on $\Lambda$. Let $v:= P(\Lambda)$. If $\Lambda$ is not a local maximum of $P$ then  there exists a sequence $x_n$ such that $d(x_n,\Lambda) \rightarrow_n 0$ and $P(x_n) > v$. Since $\Lambda$ is isolated we have $x_n \in X \setminus Z$ and $P(\varphi(x_n,t)) > P(x_n) > v$ for any $t >0$ hence $d(\varphi(x_n,t),\Lambda) \nrightarrow 0$ and $\Lambda$ is not an attractor.

Let us now prove that $(ii)$ implies $(iii)$. First we show that $\Lambda$ is contained in $NE$. Suppose that there exists $\hat{x} \in \Lambda \setminus NE$. Without loss of generality, we suppose that 
\[\hat{x}_1=0, \; \frac{\partial u_1}{\partial x_1}(\hat{x})>0.\]
Since $\frac{\partial u_1}{\partial x_1}(\hat{x}) >0$, we also have $\frac{\partial P}{\partial x_1}(\hat{x}) >0$, by definition of an LOPG. As a consequence, $\hat{x}$ is not a local maximum of $P$.

We now prove that $\Lambda$ is an attractor for the Best-Response dynamics.  $P$ is a strict Lyapunov function for the best-response dynamics\footnote{Keep in mind that this means that it is a lyapunov function with respect to NE.} and $\Lambda \subset NE$. The statement we want to prove is then a consequence of Proposition 3.25 in \cite{BHS1}. We adapt the proof in our context for convenience. First of all observe that $\Lambda$ is actually a strict local maximum of $P$: there exists an open (isolating) neighborhood $U$ of $\Lambda$ such that  $P(x) < v = P(\Lambda), \; \,  \forall x \in U \setminus \Lambda$. This is a simple consequence of the fact that $P$ is strictly increasing along any solution curve with initial conditions in $U \setminus \Lambda$. Now let $V_r := \{x \in U: P(x) > v - r\}$. Clearly $\cap_r V_r = \Lambda$. Also $\varphi(\overline{V_r},t) \subset V_r$, for $t>0$, $r$ small enough\footnote{We need to make sure that $r$ is small enough so that $\overline{V_r} = P^{-1}([v-r,v]) \subset U$}. This implies that $\Lambda = \cap_{r >0} V_r$ contains an attractor $A$. The potential being constant on $\Lambda$, $A$ cannot be strictly contained in $\Lambda$ and therefore $\Lambda$ is an attractor.   

Now clearly $(iii)$ implies $(i)$: $\Lambda = \omega_{\BR}(U)$ for some open neighborhood $U$ of $\Lambda$. Since $U \cap Z \subset NE$, $\omega_F(U) = \omega_{\BR}(U)$ and the proof is complete.
 $\; \; \blacksquare$
\vspace{.2cm}

\section{Proof of results of Section \ref{ssec:isolated}}

\subsection{Proof of Theorem \ref{th:isolated}.}

\paragraph{\underline{Proof of point $i)$}.} Pick an $\hat{x} \in OZ$ and let us fix $i \in \{1,\ldots,N \}$ such that $\hat x^i=0$ and $ \frac{\partial u^i}{\partial x^i}(\hat x) >0 $. Observe first that we can work on the event 
$\left \{ \sup_{n \to +\infty } \norm{x_n} < + \infty \right \}$ since, otherwise, there is nothing to prove. We proceed following a similar argument as in \cite{Pos97}.

Let us assume by contradiction that  $\mathbb{P} \left(\lim_n x_n =  \hat{x} \right) >0$. By continuity and from the fact that $\hat x$ is an isolated point in $OZ$, there  exists a neighborhood $\mathcal V$ of $\hat x$ such that $ \frac{\partial u^i}{\partial x^i} \geq \eta >0 $ for all $x \in \mathcal V$ and we can choose $k_* \in \mathbb N$ such that $$\mathbb{P} \left(\{\lim_n x_n =  \hat{x}  \} \cap \{  x_n \in \mathcal V, \, \text{for all } n\geq k_*\}  \right) >0.$$

Let $\tilde{U}_{n+1}^i = \epsilon_n^i \sum_{j} \epsilon_n^j \frac{\partial u_i}{\partial x_j}(x_n)$, so that 
\[x_{n+1}^i = x_n^i \left(1 + \frac{1}{n+1} \left( \frac{\partial u_i}{\partial x_i}(x_n) + \tilde{U}_{n+1}^i + \frac{\xi_{n+1}^i}{x_n^i}\right)\right).\]

Using a Taylor expansion and the fact that $\xi_n^i = \mathcal{O}(\frac{1}{n})$ and $x_n^i \geq 1/(n+1)$ for $n$ sufficiently large, we obtain that
\[\frac{1}{x_{n+1}^i} = \frac{1}{x_n^i} \left(1 - \frac{1}{n+1} \left( \frac{\partial u_i}{\partial x_i}(x_n) + \tilde{U}_{n+1}^i \right) + o\left(\frac{1}{n}\right) \right).\] 
Using that, for $n \geq k_*$, $  \frac{\partial u_i}{\partial x_i}(x_n) \geq \eta$ and $\mathbb{E}(\tilde{U}_{n+1}^i \mid \mathcal{F}_n)=0$, we obtain  
\[\mathbb{E}\left(\frac{1}{x_{n+1}^i}  - \frac{1}{x_n^i}\mid \mathcal{F}_n\right) \leq - \frac{1}{x_n^i} \cdot \frac{1}{n+1}\cdot \frac{\eta}{2} \leq 0. \]
Therefore, the random sequence $\left(1/x_n^i\right)_n$ is a positive supermartingale. It then converges almost surely to some random variable $Y$. However, on the event $\{\lim_n x_n =  \hat{x}  \}$, we have that $x_n^i$ tends to zero almost surely. These two convergence properties are in contradiction and the conclusion follows. $\; \; \blacksquare$
\vspace{.2cm}

\paragraph{\underline{Proof of point $ii)$}.}  As in the proof of Theorem \ref{th:convcompl}, we need to show that the condition \eqref{eq:noise} is verified in the unstable direction. However, we have proved in Theorem \ref{th:convcompl} part $(ii)$ that the quantity \eqref{eq:noise_unif} is always strictly positive, unless $v = 0$. $\; \; \blacksquare$  
\vspace{.2cm}

\subsection{Proof of results in Section \ref{subs:conc_games}}

\underline{\bf{Proof of Remark \ref{lm:arr_ros}:}} Following \cite{rosen1965existence}, we define $G(x,r)$ as the Jacobian matrix of $g(x,r)$, with $r_i \geq 0$. A sufficient condition for a game to belong to $\mathcal{G}^{Ros}$ is that $G(x,r) + G'(x,r)$ is negative definite, where $G'$ is the transpose of $G$. For simplicity, we set $r = \mathbf{1}$, so that $g_i(x,\mathbf{1}) = \frac{\partial u_i}{\partial x_i}$ and $G_{ij}(x,\mathbf{1}) = \frac{\partial^2 u_i}{\partial x_i \partial x_j}$, and show that games in $\mathcal{G}^{Arr}$ are such that $G(x,\mathbf{1}) + G'(x,\mathbf{1})$ is negative definite.
Define the matrices $A, B^{k}$ and $C$ as follows:\\
$A _{ii} = \frac{\partial^2 u_i}{\partial x_i^2}$ and $A_{ij} = 0$ if $i \neq j$\\
$B^{k}_{ij} = 0$ if $i = k$ or $j=k$ and $B^{k}_{ij} = \frac{\partial^2 u_k}{\partial x_i \partial x_j}$ if $i \neq k$ and $j \neq k$\\
$C_{ij} = \sum_k \frac{\partial^2 u_k}{\partial x_i \partial x_j}$\\
Then $G(x,\mathbf{1}) + G'(x,\mathbf{1}) = A(x) - \sum_k B^k(x) + C(x)$. By concavity of $u_i$ in $x_i$, $A$ is negative semi-definite and is negative definite as soon as one $u_i$ is strictly concave in $x_i$. Every $B^k$ is positive semi-definite by convexity of $u_i$ in $x_{-i}$. Finally, strictly concave-convex games are such that $\sum_k u_k(x)$ is concave in $x$, by taking $S = N$ in the definition of strictly concave-convex games. Thus $C$ is negative semi-definite.
This proves that $G(x,\mathbf{1}) + G'(x,\mathbf{1})$ is negative definite.
$\;\; \blacksquare$
\vspace{.3cm}

\underline{\bf{Proof of Theorem \ref{th:rosen}:}}
Suppose first that there is a unique Nash equilibrium. Then note that under \eqref{hy:rosen} we have, for any $x \neq \overline{x}$,
\[\langle \overline{x} - x , g(x,r)\rangle >0,\]
because 
\[\langle x - \overline{x} \mid g(\overline{x},r) \rangle = \sum_{i: \overline{x}_i = 0} r_i x_i \frac{\partial u_i}{\partial x_i}(\overline{x}) \leq 0.\]

\noindent Given an element $x \in X$, let $I(x) := \left\{i \in \mathcal{N}: \;  x_i=0 \text{ and } \, \frac{\partial u_i}{\partial x_i}(x) >0\right\}$.
Given $J \subset N$,  we call $\mathcal{G}^J$ the  $N - |J|$-player game where the set of players is $N \setminus J$ and, for any strategy profile $z \in [0 , + \infty[^{N-|J|}$, the payoff function of player $i \in \mathcal{N} \setminus J$ is  $u^J_i(z) := u_i(z,0^{|J|})$.

\begin{lemma} \label{lm:sousNash} Let $J \subset N$. There exists a unique profile $\tilde{x}^J$ with the following properties: 
\begin{itemize}
\item[$(i)$] $J \subset I(\tilde{x}^J)$,
\item[$(ii)$] $\tilde{z} := (\tilde{x}^J_i)_{i \notin J}$ is a Nash equilibrium of $\mathcal{G}^J$.
\end{itemize}
and $\tilde{x}^J \in Z(F)$. 
Moreover, if $J \subset I(\overline{x})$ then $\tilde{x}^J = \overline{x}$. If not then $\tilde{x}^J$ belongs to OZ. 
\end{lemma}

\noindent {\bfseries Proof.} Fix $J \subset N$.  The associated game $\mathcal{G}^J$ is also strictly diagonally concave. Thus it admits a unique Nash equilibrium $\tilde{z} \in [0,+\infty[^{N - |J|}$. Note that $J \subset I(\tilde{z})$ but is not necessarily equal. Now let $\tilde{x}^J := (\tilde{z},0^J)$. Clearly $\tilde{x}^J$ is the only element of $X$ satisfying both $(i)$ and $(ii)$. Let $i \notin J$. We have $\tilde{x}^J_i =\tilde{z}_i =  BR^J_i(\tilde{z}_{-i}) = BR_i(\tilde{z}_{-i},0^{J}) = BR_i(\tilde{x}^J_{-i})$. This proves that $\tilde{x}^J$ belongs to $Z(F)$.

\noindent Now suppose that $J \subset I(\overline{x})$. Then $\overline{x}$ satisfies $(i)$. Moreover for all $i \notin J$, 
\[\overline{x}_i = \BR_i (\overline{x}_{-i}) = \argmax_{x_i} u_i(x_i,\overline{x}_{-i}) = \argmax_{z_i} u^J(z_i,\overline{z}_{-i}),\]
 by definition of $u^J$ and the fact that $\overline{x}_j = 0$ for any $j \in J$. Thus $(\overline{x}_i)_{i \notin J}$ is a Nash equilibrium of $\mathcal{G}^J$ and $\tilde{x}^J = \overline{x}$. Finally if
 $J$ is not contained in $I(\hat{x})$ then $\tilde{x}^J \neq \overline{x}$ because  $\overline{x}$ does not satisfy $(i)$.  $\; \; \blacksquare$
\vspace{.3cm}

\noindent As a consequence, $\{\tilde{x}^J, J \subset I\}$ can be written as $\{\overline{x},\tilde{x}^1,...,\tilde{x}^K\}$ where all elements are distinct, and there is a natural partition of $X$:
\[X =   \left(\cup_{k=1}^K \tilde{X}^k\right) \cup \overline{X}, \; \, \text{ where } \; \tilde{X}^k := \{x \in X: \; \hat{x}^{I(x)} = \hat{x}^k\} \text{ and } \overline{X} := \left\{x \in X: \; \hat{x}^{I(x)} = \overline{x} \right\}. \]
Note that $\overline{X} = \left\{x \in X: \; I(x) \subset I(\overline{x}) \right\}$ and the sets $\overline{X}, \tilde{X}^k, k=1,...,K$ are convex. More accurately every $\overline{X}, \tilde{X}^k$ is a union of faces of $X$:  there exist $\overline{\mathcal{J}}$ and  a family $(\mathcal{J}_k)_{k=1,...,K}$ of subsets of $N$ such that:
\[\overline{X} = \cup_{J \in \mathcal{\overline{J}}} \{x \in X: I(x) = J\}\tilde{X}^k = \cup_{J \in \mathcal{J}^k} \{x \in X: I(x) = J\}.\]

Now we are ready to prove the theorem, i.e. when a game is diagonally strictly concave with unique Nash equilibrium $\overline{x}$, necessarily \[\mathbb{P}(\lim_n x_n = \overline{x}) = 1\]

First let $x \in \overline{X}$, which amounts to having $I(x) \subset I(\overline{x})$ and define, for $x \in \overline{X}$,
\[\Phi(x) = \sum_{i \in I(\overline{x})} r_i x_i + \sum_{i \notin I(\overline{x})} r_i (x_i -\overline{x}_i \log(x_i)) .\]
Then $\Phi$ is concave on $\overline{X}$ and achieves its minimum in  $\overline{x}$.
Let $\phi(t) = \Phi(\mathbf{x}(t))$, where $\mathbf{x}(t)$ is a solution of $\dot{x} = F(x)$, with $\mathbf{x}(0) \in \overline{X}$. We have

\[\frac{d}{dt} \phi(t) = \sum_{i \in \mathcal{N}}  r_i (x_i(t) -  \overline{x}_i) \frac{\partial u_i}{\partial x_i}(\mathbf{x}(t)) \leq 0,\]
with equality if and only if $x = \overline{x}$ and  $\overline{x}$ is a global attractor the flow $\Phi_{\mid \overline{X}}$.

\noindent Now suppose that $x \in \tilde{X}^k$ for a given $k \in \{1,...,K\}$. Note that $I(x) \subset I(\tilde{x}^k)$. We can then define $\Phi^k : \tilde{X}^k \rightarrow \mathbb{R}$ as the following:
\[\Phi^k(x) = \sum_{i \in I(\tilde{x}^k)} r_i x_i + \sum_{i \notin I(\tilde{x}^k) } r_i (x_i - \tilde{x}^k_i \log (x_i)).\]
Then $\Phi^k$ is again concave, with unique maximum in $x = \tilde{x}^k$ on $\tilde{X}^k$. Let $\phi(t) = \Phi(\mathbf{x}(t))$, where $\mathbf{x}(t)$ is a solution of $\dot{x} = F(x)$, with $\mathbf{x}(0) \in \tilde{X}^k$. We have

\[\frac{d}{dt} \phi(t) = \sum_{i \in \mathcal{N}}  r_i (x_i(t) -  \overline{x}_i) \frac{\partial u_i}{\partial x_i}(\mathbf{x}(t)) \leq 0,\]
with equality if and only if $x = \tilde{x}^k$. Thus $\tilde{x}^k$ is a global attractor the flow $\Phi_{\mid \tilde{X}^k}$.

As a consequence every solution curve converges to a zero of $F$, i.e either $\overline{x}$ or one of the $\tilde{x}^k$.\footnote{This is not enough to guarantee that our random process converges with probability one to one of the zeroes of the dynamics.} More precisely, $\overline{X}$ and $\tilde{X}^k$ are invariant and $\{\overline{x}\}$ (resp. $\tilde{x}^k$) is a global attractor for the flow $\phi_{\mid \overline{X}}$ (resp. $\phi_{\mid \tilde{X}^k}$); in particular, for any $x_0 \in \overline{X}$ (resp. $x_0 \in \tilde{X}^k$) then  $\lim_{t \rightarrow + \infty} \phi_t(x_0) = \overline{x}$ (resp. $\lim_{t \rightarrow + \infty} \phi_t(x_0) = \tilde{x}^k$).

A set $L$ is \emph{internally chain transitive (ICT)} for the flow $\phi_t$ if it is compact, invariant and the restriction of the flow $\phi_{\mid L}$ admits no proper attractor. Of course $L_k := \{\tilde{x}_k\}$, as well as $\overline{L} := \{\overline{x}\}$ are ICT.  

\begin{theorem}[Benaim, 1999] The limit set of $(x_n)_n$ is almost surely internally chain transitive. Moreover let $L$ be an internally chain transitive set for a flow $(\phi_t)_t$ and $A$ be an attractor with basin of attraction $\mathcal{B}(A)$. If $L \cap \mathcal{B}(A) \neq \emptyset$ then $L \subset A$. 
\end{theorem}

We now prove that the sets $L_k$ and $\overline{L}$ are the only internally chain transitive sets. This will conclude the proof. 
Note that $\overline{X}$ is an open set in $X$. To do so we first claim that it is always possible to relabel the family $(\tilde{x}^k)_{k=1,...,K}$ such that $\tilde{X}^k$ is an open set of $\cup_{l=1}^k X^l$ for $k=2,...,K$.

\noindent Let $L$ be internally chain transitive. By previous result, if $L$ intersects $\overline{X}$ then $L \subset \{\overline{x}\}$ because $\overline{X}$ is the basin of attraction of $\overline{x}$. Suppose that it is not the case. then $L \subset \cup_{k=1}^K X^k$. 
Since $X_K$ is open in $\cup_{k=1}^K X^k$, $\tilde{x}^K$ is an attractor of the flow restricted to $\cup_{k=1}^K X^k$, with basin of attraction $\tilde{X}^k$. Hence if $L \cap \tilde{X}^k \neq \emptyset$ then $L = \{\tilde{x}^k\}$. By a recursive argument, either $L = \{\overline{x}\}$ or $L = \{\tilde{x}^k\}$ for some $k$. $\; \; \blacksquare$
\vspace{.3cm}

\section{Examples}
In this section, we illustrate through two examples the importance of two conditions we have used in this paper. The first example illustrate why bipartite interaction graphs might cause some trouble, while the second shows why the assumption of symmetric externalities matters.

In every proof of non convergence, the key argument we used relied on the noise condition \eqref{eq:noise}. 
\begin{example} \label{ex_1} Consider the following $4$-player example with strategic substitutes.

\[u_1(x) = -cx_1 + b(x_1+x_2+x_4), u_2(x) = -cx_2 + b(x_2+x_1+x_3),\]
\[u_3(x) = -cx_3 + b(x_3+x_2+x_4), u_4(x) = -cx_4 + b(x_4+x_1+x_3),\]
with $b$ strictly concave and such that $b'(1)= c$. 
This is a game of strategic substitutes, with an interaction graph represented by the square in Figure \ref{fig:carre}.

\begin{figure}[h]
\begin{center}
\begin{tikzpicture}[scale=0.6, thick, every node/.style={draw,circle, minimum size=0.6cm, inner sep=0pt}]
 \node[fill=black!30!] (1) at (0,4)  {\footnotesize $1$};
 \node[fill=black!30!] (2) at (4,4)  {\footnotesize $2$};
 \node[fill=black!30!] (3) at (4,0)  {\footnotesize $3$};
 \node[fill=black!30!] (4) at (0,0)  {\footnotesize $4$};
 \draw (1) -- (2);
 \draw (2) -- (3);
 \draw (3) -- (4);
 \draw (4) -- (1);
\end{tikzpicture}
\caption{Different isolated Nash equilibria}
 \label{fig:carre}
\end{center}
\end{figure}
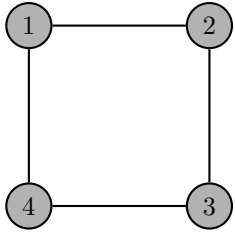

One can check that the profile $\hat{x} = (1/3,1/3,1/3,1/3)$ is a Nash equilibrium. Choosing $b$ such that $b''(1) = -3$ for simplicity, the Jacobian matrix associated to $\hat{x}$ is
\[DF(\hat{x}) = \begin{pmatrix}
-1 & -1 & 0 & -1\\
-1 & -1 & -1 & 0\\
0 & -1 & -1 & -1\\
-1 & 0 & -1 & -1
\end{pmatrix},\]
which eigenvalues are $-3,-1,-1,1$. Thus this Nash equilibrium is linearly unstable. However, the eigenspace associated to the positive eigenvalue is generated by $v=(1,-1,1,-1)$ so that, on the event $\{ \lim_n x_n = \hat{x}\}$, we have
\begin{eqnarray*}
\lim_{n \rightarrow + \infty} \mathbb{E} \left(\langle U_{n+1},v \rangle^2 \mid  \mathcal{F}_n \right) &=&  \left(\frac{\partial u_1}{\partial x_2}(\hat{x}) - \frac{\partial u_2}{\partial x_1}(\hat{x})\right)^2 \\
&+& \left(-\frac{\partial u_2}{\partial x_3}(\hat{x}) +  \frac{\partial u_3}{\partial x_2}(\hat{x})\right)^2 + \left(\frac{\partial u_3}{\partial x_4}(\hat{x}) - \frac{\partial u_4}{\partial x_3}(\hat{x})\right)^2=0
\end{eqnarray*} 
and the noise condition \eqref{eq:noise} does not hold.
\end{example}
\vspace{.3cm}

\begin{example} \label{ex_2} 
Consider the following $2$-player game with strategic complements. 
\[u_1(x_1,x_2) = - \frac{x_1^2}{2} + 2x_1 - x_1(2-x_2)^2; \; \, u_2(x_1,x_2) = - \frac{x_2^2}{2} - x_1^2(2-x_2).\]
This game has anti-symmetric externalities, since $\frac{\partial u_2}{\partial x_1}(x) = - \frac{\partial u_1}{\partial x_2}(x)$.
Now, the profile $(1,1)$ is a Nash equilibrium, and 
\[\frac{\partial^2 u_i}{\partial x_i \partial x_j}(\hat{x}) = 2, \; \, i=1,2.\]
As a consequence the Jacobian matrix associated to the dynamics $F$ is simply
\[DF(\hat{x}) = \begin{pmatrix}
-1 & 2 \\
2 & -1
\end{pmatrix},\]
which eigenvalues are $-3$ and $1$. Thus this Nash equilibrium is linearly unstable. The eigenspace associated to the positive eigenvalue is generated by $v=(1,1)$.  Thus, on the event $\{ \lim_n x_n = \hat{x}\}$, we have
\[\lim_{n \rightarrow + \infty} \mathbb{E} \left(\langle U_{n+1},v \rangle^2 \mid  \mathcal{F}_n \right) =  \left(\frac{\partial u_1}{\partial x_2}(\hat{x}) +  \frac{\partial u_2}{\partial x_1}(\hat{x})\right)^2=0\] 
and the noise condition does not hold.
\end{example}
\vspace{.2cm}

\bibliographystyle{apalike}

\end{document}